\documentclass[AMA,STIX1COL]{WileyNJD-v2}
\usepackage{subfig}
\pdfoutput=1

\articletype{Article Type}%
\received{26 April 2016}
\revised{6 June 2016}
\accepted{6 June 2016}

\ifx\optionkeymacros\undefined\else \fi

\catcode`\Œ=\active\defŒ{{\aa}}       
\catcode`\º=\active\defº{\int}        
\catcode`\=\active\def{\c c}        
\catcode`\¶=\active\def¶{\partial}    
\catcode`\Ä=\active\defÄ{\oint}       
\catcode`\Æ=\active\defÆ{\triangle}   
\catcode`\Â=\active\defÂ{\neg}        
\catcode`\µ=\active\defµ{\mu}         
\catcode`\¿=\active\def¿{{\o}}        
\catcode`\¹=\active\def¹{\pi}         
\catcode`\Ï=\active\defÏ{{\oe}}       
\catcode`\§=\active\def§{{\ss}}       
\catcode`\ =\active\def {\dagger}     
\catcode`\Ã=\active\defÃ{\sqrt}       
\catcode`\·=\active\def·{\Sigma}      
\catcode`\Å=\active\defÅ{\approx}     
\catcode`\½=\active\def½{\Omega}      
\catcode`\£=\active\def£{{\it\$}}     
\catcode`\°=\active\def°{\infty}      
\catcode`\¤=\active\def¤{{\S}}        
\catcode`\¦=\active\def¦{{\P}}        
\catcode`\¥=\active\def¥{\bullet}     
\catcode`\»=\active\def»{\leavevmode\raise.585ex\hbox{\b a}}      
\catcode`\¼=\active\def¼{\leavevmode\raise.6ex\hbox{\b o}}        
\catcode`\­=\active\def­{\not=}       
\catcode`\²=\active\def²{\leq}        
\catcode`\³=\active\def³{\geq}        
\catcode`\Ö=\active\defÖ{\div}        
\catcode`\É=\active\defÉ{{\dots}}     
\catcode`\¾=\active\def¾{{\ae}}       
\catcode`\Ç=\active\defÇ{\ll}         
\catcode`\Ò=\active\defÒ{``}          
\catcode`\Á=\active\defÁ{!`}          
\catcode`\¢=\active\def¢{\rlap/c}     
\catcode`\Ô=\active\defÔ{`}           
\catcode`\Õ=\active\defÕ{'}           

\catcode`\=\active\def{{\AA}}       
\catcode`\'=\active\def'{\c C}        
\catcode`\¯=\active\def¯{{\O}}        
\catcode`\¸=\active\def¸{\Pi}         
\catcode`\Î=\active\defÎ{{\OE}}       
\catcode`\®=\active\def®{{\AE}}       
\catcode`\×=\active\def×{\diamond}    
\catcode`\¡=\active\def¡{\accent'27}  
\catcode`\Ó=\active\defÓ{''}          
\catcode`\±=\active\def±{\pm}         
\catcode`\È=\active\defÈ{\gg}         
\catcode`\À=\active\defÀ{?`}          
\catcode`\Ð=\active\defÐ{--}          
\catcode`\Ñ=\active\defÑ{---}         

\catcode`\Š=\active\defŠ{\"a}        
\catcode`\'=\active\def'{\"e}        
\catcode`\•=\active\def•{\"{\i}}     
\catcode`\š=\active\defš{\"o}        
\catcode`\Ÿ=\active\defŸ{\"u}        
\catcode`\Ø=\active\defØ{\"y}        
\catcode`\€=\active\def€{\"A}        
\catcode`\…=\active\def…{\"O}        
\catcode`\†=\active\def†{\"U}        
\catcode`\‡=\active\def‡{\'a}        
\catcode`\Ž=\active\defŽ{\'e}        
\catcode`\'=\active\def'{\'{\i}}     
\catcode`\—=\active\def—{\'o}        
\catcode`\œ=\active\defœ{\'u}        
\catcode`\ƒ=\active\defƒ{\'E}        
\catcode`\ˆ=\active\defˆ{\`a}        
\catcode`\=\active\def{\`e}        
\catcode`\"=\active\def"{\`{\i}}     
\catcode`\˜=\active\def˜{\`o}        
\catcode`\=\active\def{\`u}        
\catcode`\Ë=\active\defË{\`A}        
\catcode`\‹=\active\def‹{\~a}        
\catcode`\–=\active\def–{\~n}        
\catcode`\›=\active\def›{\~o}        
\catcode`\Ì=\active\defÌ{\~A}        
\catcode`\"=\active\def"{\~N}        
\catcode`\Í=\active\defÍ{\~O}        
\catcode`\‰=\active\def‰{\^a}        
\catcode`\=\active\def{\^e}        
\catcode`\"=\active\def"{\^{\i}}     
\catcode`\™=\active\def™{\^o}        
\catcode`\ž=\active\defž{\^u}        

\let\optionkeymacros\null

\begin{document}

\title{A dual resolution phase-field solver for wetting of viscoelastic droplets}

\author[1]{Kazem Bazesefidpar}

\author[1,2]{Luca Brandt}

\author[1]{Outi Tammisola*}

\authormark{Bazesefidpar et al.}

\address[1]{\orgdiv{SeRC (Swedish e-Science Research Centre) and FLOW}, \orgname{Department of Engineering Mechanics, KTH}, \orgaddress{\state{SE-10044 Stockholm}, \country{Sweden}}}

\address[2]{\orgdiv{Department of Energy and Process Engineering}, \orgname{Norwegian University of Science andTechnology (NTNU)}, \orgaddress{\state{Trondheim}, \country{Norway}}}


\corres{*Outi Tammisola, FLOW and SeRC (Swedish e-Science ResearchCentre), KTH Mechanics, 10044 Stockholm, Sweden. \email{ outi@mech.kth.se}}

\presentaddress{This is sample for present address text this is sample for present address text}

\abstract[Summary]{We present a new and efficient phase-field solver for viscoelastic fluids with moving contact line based on a dual-resolution strategy. The interface between two immiscible fluids is tracked by using the Cahn-Hilliard phase-field model, and the viscoelasticity incorporated into the phase-field framework. The main challenge of this approach is to have enough resolution at the interface to approach the sharp-interface methods. The method presented here addresses this problem by solving the phase field variable on a mesh twice as fine as that used for the velocities, pressure, and polymer-stress constitutive  equations. The method is based on second-order finite  differences for the discretization of the fully coupled Navier-Stokes, polymeric constitutive and Cahn-Hilliard equations, and it is implemented in a 2D pencil-like domain decomposition to benefit from 
	existing highly-scalable parallel algorithms.
	A FFT-based solver is 
	used for  the Helmholtz and Poisson equations with different global sizes. The splitting method proposed by Dong, S. \cite{Dong2012} is used to impose the dynamic contact angle boundary conditions in the case of large density and viscosity ratios. The implementation is validated against experimental data and previous numerical studies in 2D and 3D. The results indicate that the dual-resolution approach produces nearly identical results while saving computational time for both Newtonian and viscoelastic flows in 3D.}



\maketitle


\section{Introduction}\label{sec1}

When the interface of two immiscible fluids intersects a solid wall, a moving contact line forms. The dynamics of a contact line is important 
in many industrial applications such as printing, coating and spray painting. The fluids in the above mentioned applications are usually non-Newtonian, and their rheological properties affect the contact-line dynamics significantly. For example, the elasticity of the fluid enhances the viscous bending and speeds up the contact-line motion; similarly, the fluid shear-thinning property causes the contact line to move faster \citep{Seevaratnam2007,Wei2009,Yue2012,Wang2015}. The elasticity of the droplet also plays an important role in the case of superhydrophobic  surfaces \citep{Shirtcliffe2010}, which are strongly non-wetting to Newtonian fluids. To mention an example,  \cite{Xub2018} observed a striking difference between the motion of a Boger fluid drop, which shows viscoelasticity without shear thinning, and a  Newtonian drop moving on a superhydrophobic surface. The velocity of the viscoelastic drop is notably reduced in comparison with the Newtonian drop, and complex branch-like patterns are left on the tilted superhydrophobic surface.

As regards modelling, in addition to a possibly complex rheology, multi-phase flows present interfaces between the different phases. The fluid motion in each bulk region is governed by the Navier-Stokes equations, with the dynamics of the different bulk regions connected by the boundary conditions at the interface. Different methods have been proposed for resolving the moving interface numerically, classified into two categories: interface tracking and interface capturing methods. The tracking methods specify the interface explicitly by defining meshes on the interface and the interfacial conditions as boundary conditions for the tracking equations. These methods are very accurate, but  usually need larger computational resources. Moreover, they typically fail to handle morphological changes such as breakup and coalescence \citep{Yue2004}.

The interface capturing methods use a fixed grid, and the interface is determined by the values of a scalar variable. 
The interface capturing methods include the front-tracking method \citep{Unverdi1992}, the Volume-of-Fluid (VOF) method \citep{Hirt1981,Gueyffier1999}, the Level-Set method \citep{Sussman1994}, and the Phase-Field method (PFM) \citep{Anderson1998,Abels2012}.
The boundary conditions at the interface are imposed implicitly e.g. using the continuous surface force (CSF) \citep{Brackbill1992} or ghost fluid method (GFM) \citep{Kang2000}, and a single set of governing equations can be applied over the entire domain.

In the phase-field approach of interest here, the sharp interface between two phases is replaced by a thin diffuse interface. The properties of the two fluid components and the phase-field variable $\phi$ vary smoothly within this layer, and the the interfacial mixing energy gives rise to interfacial tension \citep{Yue2004}. The phase-field method presents some interesting features over other interface capturing methods for the problems considered here: (1)  the non-Newtonian rheology can be incorporated easily owing to the energy-based variational formalism; (2) the method inherently regularizes singularities such as breakup, coalescence \citep{Yue2005}, and moving contact lines \citep{Jacqmin2000,Yue2010,Yue2012}. Indeed, the Cahn-Hilliard model has been used successfully for simulating viscoelastic droplet deformation in shear flows and the coalescence of two viscoelastic drops \citep{Yue-de2005,Yue2005}; moreover, it has been also employed to simulate the wetting and impact of non-Newtonian droplets \citep{Wang2015,Wang2017}.

Nevertheless, when solving these systems numerically,  the non-linearity induced by the free energy density function and the high-(fourth) order derivatives in the Cahn-Hilliard equation introduce a strong stability constraint on the time step \citep{Chen1998,Badalassi2003,Gao2012}. The explicit discretization of the bioharmonic operator imposes a severe time constrain dictated by $\Delta t\approx(\Delta x)^{4}$ \citep{Chen1998}, making simulations with a relatively thin interface (i.e.\ small $\Delta x$) prohibitive, even more for three-dimensional simulations. There are however strategies for treating the non-linear term in the free energy density function. The first approach is the convex splitting approach \citep{Eyre1998,Gao2014} which is unconditionally stable, with the drawback of resulting in a non-linear system. The second, widely used, approach is the stabilized scheme \citep{Shen2010}; while it is hard to design a second-order unconditionally stable scheme with this approach, it is simple and efficient. This scheme reduces the problem to two decoupled second-order equations with constant coefficients \citep{Yue2004}, which can be efficiently solved 
when a fast Poisson solver is available. The third approach is the invariant energy quadratization \citep{Guill2013}
which can be used to design unconditionally stable schemes. The main drawback of this scheme is that one needs to solve linear equations with variable coefficients. Finally, the recently proposed scalar auxiliary variable (SAV) approach, see \cite{Shen2018}, which is unconditionally stable. It gives rise to two decoupled equations with constant coefficients and usually a non-linear equation for the auxiliary variable. It should be noted that the unconditionally stable property mentioned above is only for the Cahn-Hilliard equation: when coupled with the Navier-Stokes equations with variable density and viscosity, additional  time-step restrictions may arise, depending on the discretization used for the momentum equations.

The numerical simulations of viscoelastic fluids also present important challenges: classic algorithms loose convergence when the Weissenberg number ($Wi$) exceed a threshold value, the so called High-Weissenberg number problem (HWNP) \citep{Fattal2004}; the Weissenberg number is defined as the ratio of  elastic to viscous forces ($Wi=\frac{\lambda_{H} u_{ref}}{l_{ref}}$, where $\lambda_{H}$ is the polymeric relaxation time). The log-conformation reformulation (LCR) proposed by \cite{Fattal2004, Fattal2005} can alleviate the HWNP as it  guarantees the positive definiteness of the conformation tensor during the simulation. These authors proposed to rewrite the constitutive equation for the conformation tensor by using a logarithmic transformation. Since the evolution of the stress tensor is usually steep and exponential, the polynomial interpolation is not able to approximate it adequately. The LCR has been implemented for single-phase viscoelastic  \citep{Fattal2004, Fattal2005,Hulsen2005} and later for elastoviscoplastic flows \citep{Izbassarov2021}, and to tackle the HWNP in two-phase flow solvers \citep{Lpez2019,Izbassarov2015}.

In the phase-field model, the interface can be defined as the region where the order parameter is between $-1<\phi<1$. One typically needs around 7-10 grid points
to resolve the interface; furthermore, the ratio $\frac{\lambda}{\eta}$ converges to the surface tension in the classical sense as $\eta\rightarrow0$ \citep{Yue2004}, where $\eta$ and $\lambda$ are a capillary width and mixing energy respectively. The phase-field variable is globally conserved, while mass leakage occurs between phases proportionally to $\eta$. Moreover, it has been observed that a droplet vanishes when it is smaller than a critical drop radius ($r_{c}=\frac{2^{\frac{1}{6} }}{3\pi}\eta V_{domain}$) proportional to the computational domain  and the capillary width \citep{Yue2007}. Therefore, the capillary width  
needs to be small, yet the computations affordable,
to alleviate these issues. In addition, the mobility parameter $M$ should be chosen so as to reach the so-called sharp-interface limit when the results no longer depend on the interfacial thickness \citep{Yue2010, Magaletti2013,Xu2018}. All these limitations impose requirements on the resolution to be used for two-phase flow problems; one strategy to alleviate the computational costs is adaptive mesh refinement, so to increase the resolution only where needed
\citep{Yue2006,Popinet2009} or the dual-resolution method to solve the phase variable on a finer mesh than velocities and pressure  \citep{Rudman1998,Ding2014}.

In this work, the Cahn-Hilliard phase field method is used for capturing the interface between two phases, and the Giesekus model has been employed to model the viscoelasticity and the shear-thinning of the non-Newtonian phase. We use the dual-resolution approach on a uniform staggered mesh to improve the efficiency of the computation. A twice finer mesh is used for the phase-field variable, while the other variables (e.g., velocities, pressure, and polymeric stresses) are discretized on a coarser mesh. Because the computation of the polymeric stresses (six equations in 3D) is expensive due to use of the LCR technique, we can double the resolution of the interface with a minor increase in the computational time, given that the non-Newtonian phase calculations take a large portion of the total computational time.

The paper is organised as follows: In section 2, we introduce the governing equations for an incompressible viscoelastic two-phase system. In section 3, we present the dual-resolution grid, spatial arrangement of the variables, and the coupling between the variables at different resolutions. The time and spatial discretization of the coupled equations are then reported. The validations of the numerical implementation are presented in section 4, together with a comparison between dual- and single-resolution simulations in 3D in terms of wall-clock time and scaling.

\section{Equations}\label{sec2}
\subsection{Governing equations}

We consider the moving contact line dynamics of an immiscible mixture 
of a Newtonian fluid with viscosity $\mu_{n}$ and a viscoelastic (Giesekus) fluid with solvent viscosity $\mu_{s}$, and polymeric viscosity $\mu_{p}$ of different density. We introduce the phase-field variable $\phi = \pm 1$ in the two fluids and $\phi = 0$ at the fluid/fluid interface. This problem can be modelled with the following coupled Cahn-Hilliard, Navier-Stokes system with corresponding viscoelastic constitutive model, see also \cite{Yue2004,Abels2012,Jacqmin2000,Carlson2009}.
\begin{eqnarray}
\frac{\partial{\phi}}{\partial{t}}+\nabla\cdot({{\mathbf{u}}\phi})=\nabla\cdot(M\nabla G),
\label{NS1}
\end{eqnarray}
\begin{eqnarray}
\rho(\frac{\partial{\mathbf{u}}}{\partial{t}}+({\mathbf{u}}\cdot\nabla){\mathbf{u}})+
{\mathbf{J}}\cdot\nabla{\mathbf{u}}=-\nabla{p}+\nabla\cdot{\boldsymbol\tau} 
+\nabla\cdot\mu(\nabla{{\mathbf{u}}}+
\nabla{{\mathbf{u}}^T})+G\nabla\phi+\mathbf{f_{ext}},
\label{NS2}
\end{eqnarray}
\begin{eqnarray}
\nabla\cdot{\mathbf{u}}=0,
\label{NS3}
\end{eqnarray}
\begin{eqnarray}
\boldsymbol\tau_{p}+\lambda_H(\frac{\partial{\boldsymbol\tau_{p}}}{\partial{t}}+
{\mathbf{u}}\cdot\nabla{\mathbf{\boldsymbol\tau_{p}}}-\boldsymbol\tau_{p}\nabla{{\mathbf{u}}}-
\nabla{{\mathbf{u}}^T}\boldsymbol\tau_{p})+\frac{\alpha\lambda_H}{\mu_p}(\boldsymbol\tau_{p}\cdot\boldsymbol\tau_{p})=\mu_p(\nabla{{\mathbf{u}}}+\nabla{{\mathbf{u}}^T}),
\label{NS4}
\end{eqnarray}
\begin{eqnarray}
G=\lambda(-\nabla^2{\phi}+f(\phi)), 
\label{NS5}
\end{eqnarray}
The different variables and coefficient in the above equations are as follows. $\mathbf{u}(\mathbf{x},t)$ is the velocity, $p(\mathbf{x},t)$ is the pressure, $\boldsymbol\tau(\mathbf{x},t)$ is the extra stress in the momentum equations (viscous and polymeric according to the phase),  $\boldsymbol\tau_p$ the polymer stress, $G$ is the chemical potential, $M$ the mobility parameter, $G\nabla\phi$  the surface force \citep{Jacqmin1999} and $\mathbf{f_{ext}}$  represents external forces. 
The function $f(\phi)$ in Eq. \ref{NS5} is defined as
\begin{eqnarray}
f(\phi)=\frac{1}{\eta^2}\phi(\phi^2-1),
\label{NS6}
\end{eqnarray}
where $\eta$ is the capillary width, indicating the interface thickness, $\lambda$ is the mixing energy density, and it is related to the surface tension in the sharp-interface limit by \citep{Yue2004}
\begin{eqnarray}
\lambda=\frac{3}{2\sqrt{2}}\sigma\eta,
\label{NS7}
\end{eqnarray}
where $\sigma$ is the interface surface tension. The density $\rho$ and the dynamic viscosity $\mu$ of the mixture are defined according to the indicator $\phi$
\begin{eqnarray}
\rho = \frac{(1+\phi)}{2}{\rho_1}+\frac{(1-\phi)}{2}{\rho_2},
\label{NS8}
\end{eqnarray}
\begin{eqnarray}
\mu=\frac{(1+\phi)}{2}{\mu_{s1}}+\frac{(1-\phi)}{2}{\mu_{s2}}.
\label{NS9}
\end{eqnarray}
The total viscosity of each phase is $\mu_{t}=\mu_{s}+\mu_{p}$, and $\mu_{t}=\mu_{s}$ when the phase is a Newtonian fluid. The density satisfies the following relation
\begin{eqnarray}
\frac{\partial{\rho}}{\partial{t}}+\nabla\cdot{\rho\mathbf{u}}=-\nabla\cdot\mathbf{J},
\label{NS10}
\end{eqnarray}
where $\mathbf{J}=-\frac{(\rho_1-\rho_2)}{2}M\mathbf{\nabla}{\mu}$. In Eq. \ref{NS4}, $\tau_{p}$ is the polymer stress, $\lambda_H$ is the polymer relaxation time, and $\alpha$ is the Giesekus mobility parameter.\\
There are two ways to incorporate the polymeric stresses into the phase-field framework. The first is to assume that $\mu_{p}$, $\lambda_H$, and $\alpha$, are a linear function of the phase-field variable 
\begin{eqnarray}
\theta=\frac{(1+\phi)}{2}{\theta_{1}}+\frac{(1-\phi)}{2}{\theta_{2}},
\label{NS11}
\end{eqnarray}
where $\theta$ is a generic non-Newtonian property, and solve the Eq. \ref{NS4} in the domain of interest. 
In this approach $\boldsymbol\tau=\boldsymbol\tau_{p}$ in Eq. \ref{NS2}. 
The second method is to assume $\mu_{p}$, $\lambda_H$, and $\alpha$ are constant and solve Eq. \ref{NS4} only in the region occupied by the non-Newtonian fluid. 
This reduces to defining $\boldsymbol\tau$ in Eq. \ref{NS2}  as \citep{Yue2005}
\begin{eqnarray}
\boldsymbol\tau=\frac{(1\pm\phi)}{2}\boldsymbol\tau_{p}, 
\label{NS12}
\end{eqnarray}
Both approaches have been implemented and tested successfully in the code. The results reported here are obtained by using the second approach.\\ 
The following boundary conditions are imposed on a solid substrate
\begin{eqnarray}
\boldsymbol{u}=\mathbf{u_w},
\label{NS13}
\end{eqnarray}
\begin{eqnarray}
{\mathbf{n}}\cdot\nabla{G}=0,
\label{NS14}
\end{eqnarray}
\begin{eqnarray}
-D_{w}(\frac{\partial{\phi}}{\partial t}+({\mathbf{u}}\cdot\nabla){\phi})={\mathbf{n}}\cdot\nabla{\phi}+\frac{1}{\lambda}f_w^\prime(\phi),
\label{NS15}
\end{eqnarray}
\begin{eqnarray}
f_{w}(\phi)=\sigma\cos(\theta_s)\frac{\phi(\phi^2-3)}{4}+\frac{(\sigma_1w+\sigma_2w)}{2},
\label{NS16}
\end{eqnarray}
where $u_w$ is the wall velocity, $\mathbf{n}$ is the outward pointing normal vector to the boundary, and $f_{w}(\phi)$ is the wall energy . Eq.\ \ref{NS13} is the no-slip boundary condition and the Eq. \ref{NS14} imposes zero flux across the solid boundary, impermeability. 
The dynamic contact line condition in Eq.\ \ref{NS15} allows us to model the relaxation of the dynamic angle to the equilibrium angle $\theta_s$ defined in Eq.\ \ref{NS16} \citep{Jacqmin2000,Carlson2009}. Relaxation at the wall tends to  retard the motion of the contact line. If the phenomenological parameter $D_{w}=0$, the fluid layer is at equilibrium with the solid substrate $\lambda{\mathbf{n}}\cdot\nabla{\phi}+f_w^\prime(\phi)=0$.

\subsection{Log conformation formulation}

The log-conformation method proposed by \cite{Fattal2004,Fattal2005} is used to relieve the High-Weissenberg number problem: the logarithm of the conformation tensor $\boldsymbol{\Theta}=\log{\boldsymbol{c}}$ is advanced in time instead of the polymer stresses, see Eq.\ \ref{NS4}. The relationship between polymer stress $\mathbf{\boldsymbol\tau_{p}}$ and the conformation tensor $\mathbf{\boldsymbol{c}}$ for the Oldroyd-B and Giesekus models is given by 
\begin{eqnarray}
\mathbf{\boldsymbol\tau_{p}}=\frac{\mu_{p}}{\lambda_{H}}(\mathbf{\boldsymbol{c}}-\mathbf{\boldsymbol{I}}),
\label{NS17}
\end{eqnarray}
Rewriting the constitutive equation Eq. \ref{NS4} in terms of conformation tensor yields
\begin{eqnarray}
\frac{\partial{\mathbf{c}}}{\partial{t}}+
{\mathbf{u}}\cdot\nabla{\mathbf{\boldsymbol{c}}}-\boldsymbol{c}\nabla{{\mathbf{u}}}-
\nabla{{\mathbf{u}}^T}\boldsymbol{c}=\frac{1}{\lambda_H}[\boldsymbol{I}+\alpha(\boldsymbol{c}-\boldsymbol{I})]\cdot(\boldsymbol{c}-\boldsymbol{I}),
\label{NS18}
\end{eqnarray}
Since $\boldsymbol{c}$ is symmetric and positive-definite, it is diagonalizable
\begin{eqnarray}
\boldsymbol{c}=\boldsymbol{R}\cdot\boldsymbol{\Lambda}\cdot\boldsymbol{R}^{T},
\label{NS19}
\end{eqnarray}
where $\boldsymbol{R}$ is the orthogonal matrix formed by the eigenvectors of $\boldsymbol{c}$ and $\boldsymbol{\Lambda}$ is the diagonal matrix defined by the corresponding eigenvalues. The log-conformation tensor $\Theta$ is defined as
\begin{eqnarray}
\boldsymbol{\Theta}=\log{\boldsymbol{c}}=\boldsymbol{R}\cdot\log{\boldsymbol{\Lambda}}\cdot\boldsymbol{R}^{T},
\label{NS20}
\end{eqnarray}
Since the velocity field is divergence-free, the velocity gradient can be decomposed as
\begin{eqnarray}
\nabla{{\mathbf{u}}}^{T}=\boldsymbol{\Omega}+\boldsymbol{B}+\boldsymbol{N}\cdot\boldsymbol{A}^{-1},
\label{NS21}
\end{eqnarray}
where $\boldsymbol{\Omega}$ and $\boldsymbol{N}$ are antisymmetric and $\boldsymbol{B}$ is symmetric and commutes with $\boldsymbol{c}$. Substituting the above decomposition into Eq.\ \ref{NS18} and using Eq.\ \ref{NS20} the equation for $\boldsymbol{\Theta}$ becomes
\begin{eqnarray}
\frac{\partial{\mathbf{\Theta}}}{\partial{t}}+
\nabla\cdot{\mathbf{(\mathbf{u}\boldsymbol{\Theta}})}-(\mathbf{\Omega}\cdot\mathbf{\Theta}-\mathbf{\Theta}\cdot\mathbf{\Omega})-
2\boldsymbol{B}=
\frac{1}{\lambda_H}(e^{-\mathbf{\Theta}}\cdot[\boldsymbol{I}+\alpha(e^{\mathbf{\Theta}}-\boldsymbol{I})]\cdot(e^{\mathbf{\Theta}}-\boldsymbol{I})),
\label{NS22}
\end{eqnarray}
where
\begin{equation}
\boldsymbol{M} 
= \boldsymbol{R}^{T}\cdot(\nabla{{\mathbf{u}}})^{T}\cdot\boldsymbol{R}
\label{NS23}
\end{equation}

and 
\begin{eqnarray}
\begin{aligned}
\boldsymbol{\Omega} & = \boldsymbol{R}
\begin{bmatrix}
0 & \omega_{xy} & \omega_{xz}\\
-\omega_{xy} & 0 & \omega_{yz}\\
-\omega_{xz} & -\omega_{yz} & 0
\end{bmatrix}
\boldsymbol{R}^{T}\\
\boldsymbol{B} &= \boldsymbol{R}
\begin{bmatrix}
m_{xx} & 0 & 0\\
0 & m_{yy} & 0\\
0 & 0 & m_{zz}
\end{bmatrix}
\boldsymbol{R}^{T}
\end{aligned}
\label{NS24}
\end{eqnarray}
where the element $\omega_{ij}=\frac{\lambda_{j}m_{ij}+\lambda_{i}m_{ji}}{\lambda_{j}-\lambda_{i}}$. In the case of $\mathbf{\boldsymbol\tau_{p}}=0$, we set $\boldsymbol{\Omega}=0$ and $\boldsymbol{B}=\frac{1}{2}[\nabla{{\mathbf{u}}}^{T}+\nabla{{\mathbf{u}}}]$.

\section{Numerical method}\label{sec3}

\subsection{Dual-grid arrangement}

\begin{figure}[tbp]
	\centering
	\includegraphics[width=0.4\textwidth]{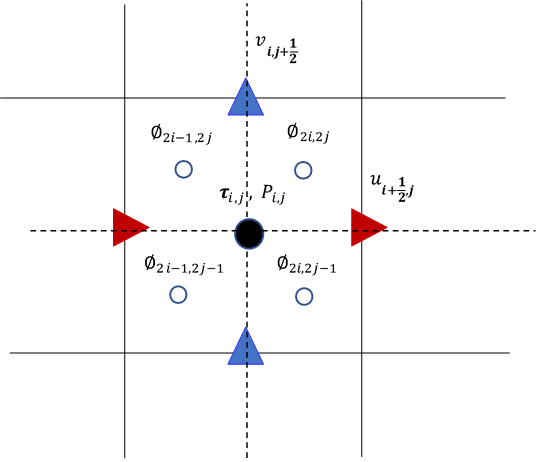}
	\caption{A 2D sketch of the dual-resolution grid: The pressure and polymer stresses are defined at the cell center (filled circle), velocity components at the cell faces (triangles), and the phase-field variable at double resolution on 4 subcells, (empty circles).}
	\label{fig:fig1}
\end{figure}

A dual-resolution grid is used to store and solve the variables at different resolutions. The velocities, pressure, and polymeric stresses are stored on a coarse grid, while the phase field variable is defined and solved on a finer grid. A uniform staggered Cartesian grid is used for the coarse mesh, with the pressure and the stresses evaluated at the cell center, and the velocity components at the cell faces. The phase-field grid is constructed by halving the pressure control volume in each direction and placing the phase-field nodes at the center of the resultant control volumes. A 2D  sketch of the dual-resolution grid and the corresponding cell indexing are presented in Fig.\ref{fig:fig1}. The relations between the different variables on the coarse mesh and fine mesh are given below in 2D; these relations can be straightforwardly extended to 3D. 

The velocity components are needed on the cell faces of the fine grid to advance the phase-field variable in time; it is also necessary to make sure that the velocities interpolated  from the coarse to the  fine mesh are divergence-free. Here, we use the same first-order interpolation for the velocity field used in \cite{Rudman1998} to guarantee that the interpolated velocity field is divergence-free. The relations between the fine-grid and coarse-grid velocities are given by
\begin{align}
\label{dual1}
\begin{split}
u_{2i+\frac{1}{2},2j-1}&=u_{2i+\frac{1}{2},2j}=u_{i+\frac{1}{2},j} 
\\
u_{2i-\frac{1}{2},2j-1}&=u_{2i-\frac{1}{2},2j}=\frac{u_{i+\frac{1}{2},j}+u_{i-\frac{1}{2},j}}{2}
\\
u_{2i-\frac{3}{2},2j-1}&=u_{2i-\frac{3}{2},2j}=u_{i-\frac{1}{2},j}
\\
v_{2i-1,2j+\frac{1}{2}}&=v_{2i,2j+\frac{1}{2}}=v_{i,j+\frac{1}{2}}
\\
v_{2i-1,2j-\frac{1}{2}}&=v_{2i,2j-\frac{1}{2}}=\frac{v_{i,j+\frac{1}{2}}+v_{i,j-\frac{1}{2}}}{2}
\\
v_{2i-1,2j-\frac{3}{2}}&=v_{2i,2j-\frac{3}{2}}=v_{i,j-\frac{1}{2}},
\end{split}
\end{align}

The polymeric viscosity, polymer relaxation time, and Giesekus mobility are needed at the cell center of the coarser mesh for solving the polymeric constitutive equations, whereas the solvent viscosity, density, and chemical potential are required on the cell faces to advance the velocity field $\mathbf{u}$. All these variables are a function of the phase-field variable $\phi$ except the chemical potential, so they can be computed at the location of interest once $\phi$ is  known.
The chemical potential and phase-field variable are computed at nodes of the coarse grid by averaging over the neighboring cell center values of the fine grid as in \cite{Rudman1998},
\begin{align}
\label{dual2}
\begin{split}
g_{i,j}&=\frac{g_{2i,2j}+g_{2i-1,2j}+g_{2i-1,2j-1}+g_{2i,2j-1}}{4} 
\\
g_{i+\frac{1}{2},j}&=\frac{g_{2i,2j}+g_{2i+1,2j}+g_{2i+1,2j-1}+g_{2i,2j-1}}{4}
\\
g_{i,j+\frac{1}{2}}&=\frac{g_{2i,2j}+g_{2i,2j+1}+g_{2i+1,2j+1}+g_{2i-1,2j}}{4},
\end{split}
\end{align}
where the function g represents $\phi$ or G (chemical potential). Similarly, the fluid properties at nodes of the coarse grid are then calculated by
\begin{align}
\label{dual3}
\begin{split}
\theta_{i,j}&=\frac{(1+\phi_{i,j})}{2}{\theta_{1}}+\frac{(1-\phi_{i,j})}{2}{\theta_{2}} 
\\
\theta_{i+\frac{1}{2},j}&=\frac{(1+\phi_{i+\frac{1}{2},j})}{2}{\theta_{1}}+\frac{(1-\phi_{i+\frac{1}{2},j})}{2}{\theta_{2}}
\\
\theta_{i,j+\frac{1}{2}}&=\frac{(1+\phi_{i,j+\frac{1}{2}})}{2}{\theta_{1}}+\frac{(1-\phi_{i,j+\frac{1}{2}})}{2}{\theta_{2}},
\end{split}
\end{align}
where $\theta$ is a generic physical property.

\subsection{Numerical algorithm and discretization in time and space}
The governing equations are discretized by using a finite difference method, both for the single- and dual-resolution solutions. 
\subsubsection{Cahn-Hilliard equation}
To benefit from the fast FFT solver, we follow the temporal discretization scheme in \cite{Dong2012} to discretize Eq.\ (\ref{NS1}) with the boundary condition (\ref{NS15}).
Using the stabilized scheme gives rise to two fully decoupled Helmholtz equations with constant coefficients. The prolongation of the velocity field, Eq. (\ref{dual1}), is needed before proceeding with the time discretization in the case of dual-resolution. First, the Cahn-Hilliard Eq.\ (\ref{NS1}) is discretized in time without considering the boundary conditions as follows
\begin{eqnarray}
\frac{\gamma_{0}\phi^{n+1}-\hat{\phi}}{\Delta{t}}+\nabla\cdot({{\mathbf{u}^{*,n+1}}\phi^{*,n+1}})=
-M\lambda\nabla^{2}[\nabla^{2}\phi^{n+1}-\frac{S}{\eta^{2}}(\phi^{n+1}-\phi^{*,n+1})-f(\phi^{*,n+1})],
\label{NS25}
\end{eqnarray}
Eq. (\ref{NS25}) can be transformed into two decoupled Helmholtz-type equations
\begin{eqnarray}
\nabla^{2}\Psi^{n+1}-(\alpha+\frac{S}{\eta^{2}})\Psi^{n+1}=Q
\label{NS26}
\end{eqnarray}
\begin{eqnarray}
\nabla^{2}\phi^{n+1}+\alpha\phi^{n+1}=\Psi^{n+1}
\label{NS27}
\end{eqnarray}
where
\begin{eqnarray}
Q=\frac{1}{M\lambda}(\frac{\hat{\phi}}{\Delta{t}}-\nabla\cdot({{\mathbf{u}^{*,n+1}}\phi^{*,n+1}}))+\nabla^{2}[f(\phi^{*,n+1})-\frac{S}{\eta^{2}}\phi^{*,n+1}].
\label{NS28}
\end{eqnarray}
The coefficients $\alpha$ and $S$ are defined as
\begin{eqnarray}
\alpha=-\frac{S}{2\eta^{2}}(1+\sqrt{1-\frac{4\gamma_{0}\eta^{4}}{M\lambda\Delta{t}S^2}})
\label{NS29}
\end{eqnarray}
and
\begin{eqnarray}
\frac{S}{\eta^{2}}\geq\sqrt{\frac{4\gamma_{0}}{M\lambda\Delta{t}}}.
\label{NS30}
\end{eqnarray}
The second-order backward approximation of any arbitrary variable $g$ is defined as 
\begin{eqnarray}
\frac{\partial{g}}{\partial t}\approx\frac{\gamma_{0}g^{n+1}-\hat{g}}{\Delta{t}}, \qquad \gamma_{0} = \frac{3}{2}, \qquad \hat{g}=2g^{n}-\frac{1}{2}g^{n-1}
\label{NS31}
\end{eqnarray}
where $g^{*,n+1}$ represents the second-order explicit approximation of $g^{n+1}$ given by
\begin{eqnarray}
g^{*,n+1}=2g^{n}-g^{n-1}
\label{NS32}
\end{eqnarray}
Equations (\ref{NS26}) and (\ref{NS27}) need to be supplemented with boundary conditions in three directions $x_{i}\equiv\{x,y,z\}$. We impose the dynamic contact angle boundary condition in the $z$-direction, and $D_{w}=0$ in the $x-$ and $y-$ directions. The latter is formally a static contact angle boundary condition; periodic and Neumann boundary conditions, specified by ${\mathbf{n}}\cdot\nabla{\phi}=0$, are also obtained from the same condition by assigning $\theta_{s}=90$.
The boundary condition in (\ref{NS33}), which enforces the global conservation of mass, and in (\ref{NS34}), which imposes the contact angle boundary condition, are given in discrete form as
\begin{eqnarray}
{\mathbf{n}}\cdot\nabla{\Psi^{n+1}}={\mathbf{n}}\cdot\nabla({f(\phi^{*,n+1})-\frac{S}{\eta^{2}}\phi^{*,n+1}})+(\alpha+\frac{S}{\eta^{2}}){\mathbf{n}}\cdot\nabla{\phi^{n+1}},
\label{NS33}
\end{eqnarray}
\begin{eqnarray}
{\mathbf{n}}\cdot\nabla{\phi^{n+1}}+D_{w}\frac{\partial{\phi}}{\partial t}^{n+1}=-\frac{1}{\lambda}f_w^\prime(\phi^{*,n+1})-D_{w}\mathbf{u}^{*,n+1}\cdot\nabla{\phi^{*,n+1}}.
\label{NS34}
\end{eqnarray}

We use second-order central difference to approximate spatial derivatives, except for the advection terms in Eq.\ (\ref{NS25}) for which the fifth-order WENO-Z is used \citep{Borges2008}. The two decoupled constant-coefficient Helmholtz equations, (\ref{NS26}) and (\ref{NS27}), with boundary conditions  (\ref{NS33}) and (\ref{NS34}), are implemented and solved by the FFT solver of the CANS code \citep{Costa2018}.  

\subsubsection{Navier-Stokes equations}
The density, viscosity, phase-field variable, and chemical potential are calculated by using the restriction operations (\ref{dual2},\ref{dual3}) in the case of dual-resolution. One needs to consider the stiffness of the viscous term at low Reynolds number and the resultant Poisson equation with variable coefficient in the pressure correction step in order to devise an efficient scheme to solve the Eq.\ (\ref{NS2}) with constrain (\ref{NS3}) in the presence of variable density and viscosity. To get the most out of the fast FFT solver, we add/subtract $\frac{\rho}{\rho_{0}}\nabla{p}$ and $\nu_{m}\rho\nabla^{2}\mathbf{u}$ to the Eq. (\ref{NS2}) \citep{Dong2012b} and use the second-order backward difference scheme for the time discretization. Assuming the viscosity, density, chemical potential, and phase-field variable known at time $n+1$, we write

\begin{equation} \label{NS35}
\begin{split}
\frac{\gamma_{0}\mathbf{u}^{n+1}-\hat{\mathbf{u}}}{\Delta{t}}=-\nabla\cdot({{\mathbf{u}^{*,n+1}}\mathbf{u}^{*,n+1}})-
\frac{1}{\rho^{n+1}}{\mathbf{J}^{n+1}}\cdot\nabla{\mathbf{u}}^{n+1}-
\frac{1}{\rho_{0}}\nabla{p^{n+1}}-(\frac{1}{\rho^{n+1}}-\frac{1}{\rho_{0}})\nabla{p^{*,n+1}}+\nu_{m}\nabla^{2}\mathbf{u}^{n+1}\\-\nu_{m}\nabla^{2}\mathbf{u}^{*,n+1}+\frac{1}{\rho^{n+1}}\nabla\cdot{\boldsymbol\tau^{*,n+1}}+
\frac{1}{\rho^{n+1}}\nabla\cdot\mu^{n+1}(\nabla{{\mathbf{u}}^{*,n+1}}+
\nabla{{\mathbf{u}^{*,n+1}}^T})+\frac{1}{\rho^{n+1}}G^{n+1}\nabla\phi^{n+1}
\end{split}
\end{equation}

In our solution algorithm, we first find an intermediate velocity field in step 1, the incremental pressure  $p^{\prime}$  in step 2 \citep{Guermond2006,Guermond2009} and finally, the corrected velocity field and pressure in step 3.
The prediction velocity, pressure, and divergence-free velocities are obtained  as follows:\\
{\bf Step 1} : solve for the prediction velocity $\tilde{\mathbf{u}}^{n+1}$

\begin{equation} \label{NS36}
\begin{split}
\nu_{m}\nabla^{2}\tilde{\mathbf{u}}^{n+1}-\frac{\gamma_{0}}{\Delta{t}}\tilde{\mathbf{u}}^{n+1}=\nabla\cdot({{\mathbf{u}^{*,n+1}}\mathbf{u}^{*,n+1}})+
\frac{1}{\rho^{n+1}}{\mathbf{J}^{n+1}}\cdot\nabla{\mathbf{u}}^{n+1}+
\frac{1}{\rho_{0}}\nabla{p^{n}}+(\frac{1}{\rho^{n+1}}-\frac{1}{\rho_{0}})\nabla{p^{*,n+1}}+\\\nu_{m}\nabla^{2}\mathbf{u}^{*,n+1}-\frac{1}{\rho^{n+1}}\nabla\cdot{\boldsymbol\tau^{*,n+1}}-
\frac{1}{\rho^{n+1}}\nabla\cdot\mu^{n+1}(\nabla{{\mathbf{u}}^{*,n+1}}+
\nabla{{\mathbf{u}^{*,n+1}}^T})-\frac{1}{\rho^{n+1}}G^{n+1}\nabla\phi^{n+1}
\end{split}
\end{equation}
{\bf Step 2} : solve for the incremental pressure $p^{\prime}$ by enforcing $\nabla\cdot{\mathbf{u}^{n+1}}=0$
\begin{equation}
\begin{cases}
\nabla^{2}{p^{\prime}}=\frac{\Delta{t}}{\gamma_{0}\rho_{0}}\nabla\cdot{\tilde{\mathbf{u}}^{n+1}}\\
\frac{\partial{p^{\prime}}}{\partial{n}}=0, \qquad on \qquad \partial{\Omega}
\end{cases}  
\label{NS37}
\end{equation}
{\bf Step 3} : the predicted velocity is corrected and  pressure is updated
\begin{eqnarray}
\mathbf{u}^{n+1}=\tilde{\mathbf{u}}^{n+1}-\frac{\Delta{t}}{\gamma_{0}\rho_{0}}\nabla{p^{\prime}},
\label{NS38}
\end{eqnarray}
\begin{eqnarray}
p^{n+1}=p^{n}+p^{\prime}.
\label{NS39}
\end{eqnarray}
In theses equations, $\rho_{0}$ and $\nu_{m}$ are constant and chosen as $\rho_{0}=min(\rho_{1},\rho_{2})$ and $\nu_{m}\geq\frac{1}{2}\frac{max(\mu_{1},\mu_{2})}{min(\rho_{1},\rho_{2})}$. The second-order central difference scheme is used for the spatial discretization of the above equations. It should be noted that both explicit and implicit treatments of the viscous term are implemented and tested in the solver, while the implicit version of the algorithm is presented here. The viscous term is discretized by the second-order backward difference in time in the explicit version. 
The three decoupled constant-coefficient Helmholtz equations (\ref{NS36}) and one constant-coefficient Poisson equation (\ref{NS37}) for the pressure are solved with the FFT solver in \cite{Costa2018}.

\subsubsection{The polymeric stresses}
After updating the  phase-field, velocity, and pressure fields, the polymeric stress tensor remains to be updated. The polymeric viscosity, polymer relaxation time, Giesekus mobility, and phase-field variable are computed at the cell center of the coarse grid with the restriction operations (\ref{dual2},\ref{dual3}) in the case of dual-resolution. The second-order central difference scheme is used to approximate spatial derivatives, with the exception of the convective term approximated by the fifth-order WENO-Z scheme \citep{Borges2008}. For the sake of clarity, we rewrite the Eq.\ (\ref{NS22}) in the following from
\begin{eqnarray}
\frac{\partial{\mathbf{\Theta}}}{\partial{t}}={\mathbf{RH}},
\label{NS40}
\end{eqnarray}
with
\begin{eqnarray}
\mathbf{RH}=-\nabla\cdot{\mathbf{(\mathbf{u}\boldsymbol{\Theta}})}+(\mathbf{\Omega}\cdot\mathbf{\Theta}-\mathbf{\Theta}\cdot\mathbf{\Omega})+
2\boldsymbol{B}+
\frac{1}{\lambda_H}(e^{-\mathbf{\Theta}}\cdot[\boldsymbol{I}+\alpha(e^{\mathbf{\Theta}}-\boldsymbol{I})]\cdot(e^{\mathbf{\Theta}}-\boldsymbol{I})).
\label{NS41}
\end{eqnarray}
A second-order total variation diminishing (TVD) Runge-Kutta method \citep{Gottlieb1998} is used for the the temporal discretization of Eq.\ (\ref{NS40}). We advance the logarithm of the conformation tensor with the known velocity field at time $n+1$, $\mathbf{u}^{n+1}$ and the phase-field variable $\phi^{n+1}$ as follows:

\vspace{0.2cm}\noindent
{\bf step 1} : The gradient velocity $\nabla{{\mathbf{u}}}^{n+1}$ is decomposed using $\boldsymbol{R}^{n}$ and $\boldsymbol{\Lambda}^{n}$ from the previous time step, see Eq.\ (\ref{NS23}), to compute $\boldsymbol{\Omega}$ and $\boldsymbol{B}$ as defined in Eq.\ (\ref{NS24}).
\begin{eqnarray}
\mathbf{\Theta}^{(1)} = \mathbf{\Theta}^{n} + \Delta{t}\mathbf{RH}^{n}.
\label{NS42}
\end{eqnarray}
{\bf step 2} : The log of conformation tensor is diagonalised and $\boldsymbol{R}^{(1)}$ and $\boldsymbol{\Lambda}^{(1)}$ computed as
\begin{eqnarray}
\boldsymbol{\Theta}^{(1)}=\boldsymbol{R}^{(1)}\cdot{\boldsymbol{\Lambda}_{{\Theta}}^{(1)}}\cdot{\boldsymbol{R}^{(1)}}^{T}, \qquad \boldsymbol{\Lambda}^{(1)} = e^{\boldsymbol{\Lambda}_{{\Theta}}^{(1)}}.
\label{NS43}
\end{eqnarray}
{\bf step 3} :  $\mathbf{\Theta}^{n+1}$ is computed using $\boldsymbol{R}^{(1)}$ and $\boldsymbol{\Lambda}^{(1)}$ to calculate $\boldsymbol{\Omega}^{(1)}$ and $\boldsymbol{B}^{(1)}$, see Eq.\ (\ref{NS24}),
\begin{eqnarray}
\mathbf{\Theta}^{n+1} = \frac{1}{2}\mathbf{\Theta}^{n} + \frac{1}{2}(\mathbf{\Theta}^{(1)}+\Delta{t}\mathbf{RH}^{(1)}).
\label{NS44}
\end{eqnarray}
{\bf step 4} : $\mathbf{\Theta}^{n+1}$ is  diagonalised, and $\boldsymbol{R}^{n+1}$ and $\boldsymbol{\Lambda}^{n+1}$ are stored for the next time step 
\begin{eqnarray}
\boldsymbol{\Theta}^{n+1}=\boldsymbol{R}^{n+1}\cdot{\boldsymbol{\Lambda}_{{\Theta}}^{n+1}}\cdot{\boldsymbol{R}^{n+1}}^{T}, \qquad \boldsymbol{\Lambda}^{n+1} = e^{\boldsymbol{\Lambda}_{{\Theta}}^{n+1}},
\label{NS45}
\end{eqnarray}
{\bf step 5} : $\mathbf{\boldsymbol\tau_{p}}^{n+1}$ is computed according to
\begin{eqnarray}
\boldsymbol{c}^{n+1}=\boldsymbol{R}^{n+1}\cdot\boldsymbol{\Lambda}^{n+1}\cdot{\boldsymbol{R}^{n+1}}^{T}, \qquad \mathbf{\boldsymbol\tau_{p}}^{n+1}=\frac{\mu_{p}^{n+1}}{\lambda_{H}^{n+1}}(\mathbf{\boldsymbol{c}}^{n+1}-\mathbf{\boldsymbol{I}}).
\label{NS46}
\end{eqnarray}
Boundary conditions should be considered just at the inflow for the polymeric constitutive equations owing to their hyperbolic nature \citep{Van1988,Figueiredo2016}, hence homogeneous Neumann boundary conditions are imposed for $\boldsymbol\tau_{p}$ and $\boldsymbol{\Theta}$ at inflow boundaries.
\section{Validations}\label{sec4}
In this Section, we present the validations and performance tests of the viscoelastic two-phase flow solver with the Cahn-Hilliard model with moving contact line against several benchmarks and different previous numerical studies in 2D and 3D.
\begin{figure}[tbp]
	\centering
	\includegraphics[width=0.5\textwidth]{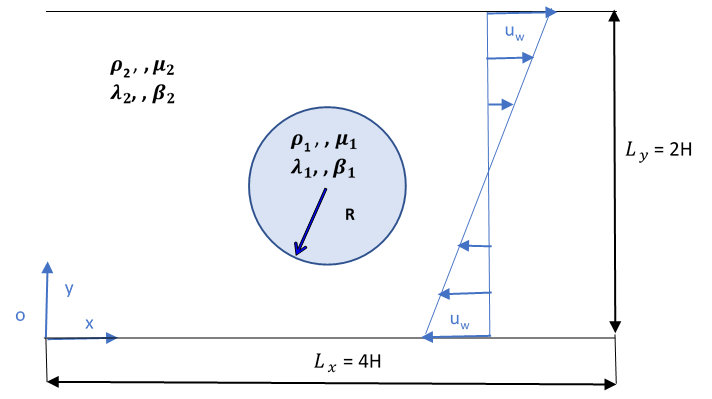}
	\caption{Initial configuration for the case of a viscoelastic droplet  in the Couette flow.}
	\label{fig:fig2}
\end{figure}

\begin{figure}[tbp]{}
	\subfloat[]{\includegraphics[width = 2.5in]{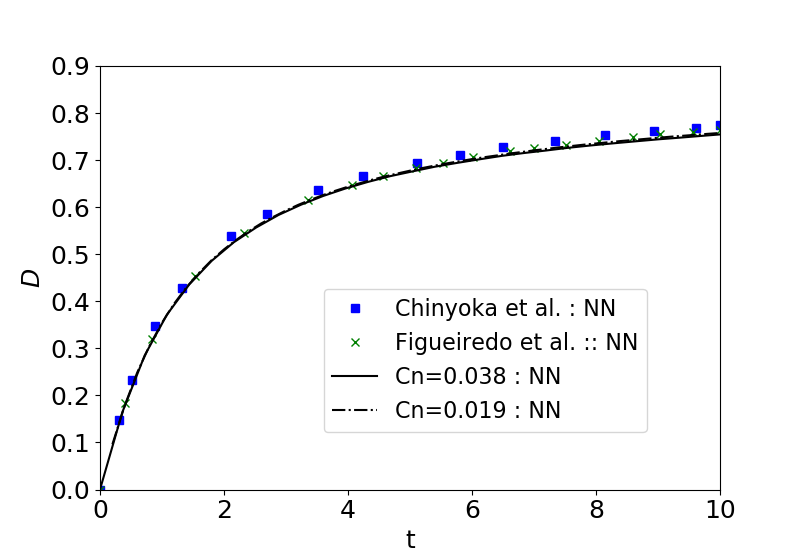}} 
	\centering
	\subfloat[]{\includegraphics[width = 2.5in]{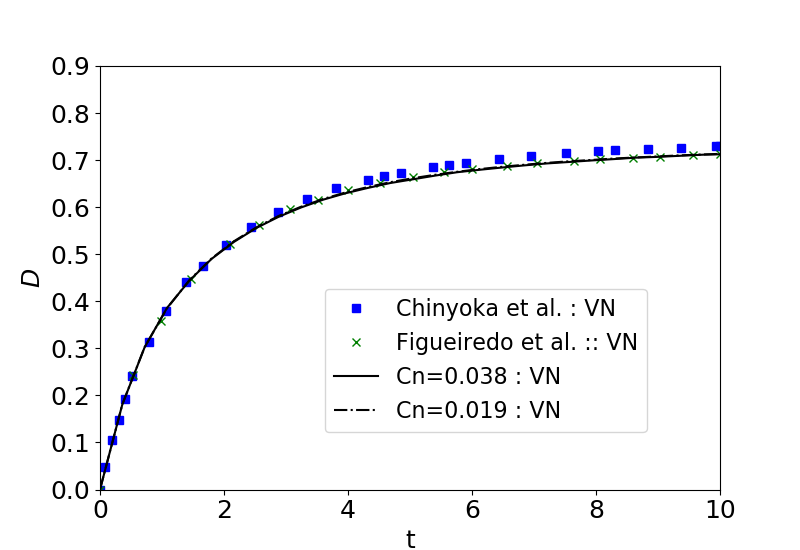}}\\
	\centering
	\subfloat[]{\includegraphics[width = 3in]{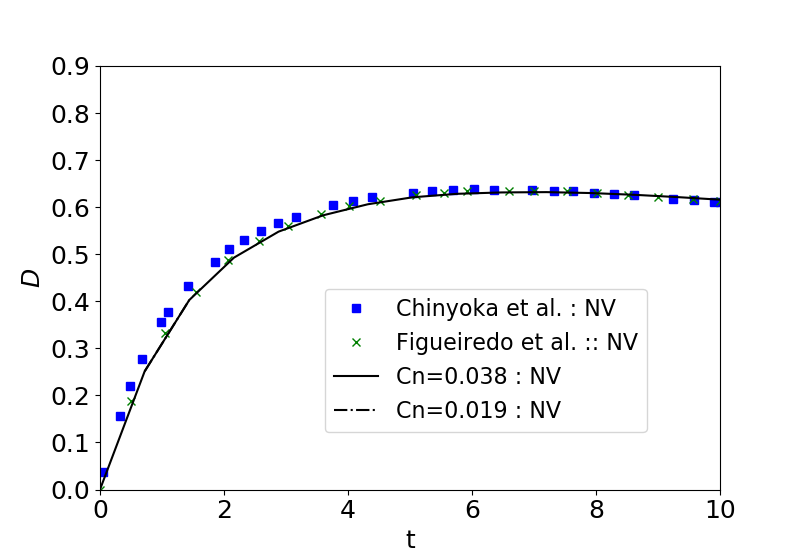}}
\caption{Evolution of the droplet deformation parameter $D$ for (a) a Newtonian droplet in a Newtonian fluid ($NN$); (b) Newtonian droplet in a viscoelastic fluid ($VN$) (c) Viscoelastic droplet in a Newtonian fluid ($NV$). The simulation data are compared with the results in  \cite{Chinyoka2005,Figueiredo2016} (symbols).}
\label{fig:fig3}
\end{figure}

\subsection{Droplet deformation in Couette flow}
We start by validating our numerical implementation with the case of two-dimensional (2D) droplet deformation under a constant shear rate.  
First, we will validate the single-resolution solver; we will examine the accuracy and efficiency of the dual-resolution method in the next section.
The configuration used for this case is presented in Fig.\ \ref{fig:fig2}. A droplet is placed between two parallel plates, where the bottom wall moves with the velocity $-u_{w}$ and the top wall with the velocity $u_{w}$. The droplet initial radius is $\frac{H}{R}=4$ and the shear rate $\dot{\gamma}=\frac{u_{w}}{H}$. The constitutive model  is the Oldroyd-B model.\\
This test case is defined by the following dimensionless numbers: the Reynolds number, $Re=\frac{\rho_{2}\dot{\gamma}R^{2}}{\mu_{2}}$, the ratio between
inertial and viscous forces; the Capillary number, $Ca=\frac{\mu_{2}\dot{\gamma}R}{\sigma}$, the ratio between viscous and surface tension forces; $Wi=\dot{\gamma}\lambda_{H}$ the ratio between elastic and viscous forces; the Peclet number, $Pe=\frac{2\sqrt{2}\dot{\gamma}R^{2}\eta}{3M\sigma}$, which is the ratio between the advection and diffusion; the Cahn number, $Cn=\frac{\eta}{R}$, which represents the ratio between the interface width and the characteristic length scale; the relaxation time ratio, $\beta=\frac{\mu_{s}}{\mu_{s}+\mu_{p}}$, which is the ratio between the polymeric viscosity and total viscosity; the viscosity ratio, $\lambda_{\mu}=\frac{\mu_{2}}{\mu_{1}}$, which is the ratio between ambient viscosity to droplet viscosity; and finally, density ratio, $\lambda_{\rho}=\frac{\rho_{2}}{\rho_{1}}$, the ratio between ambient density to droplet density.
The droplet deformation will be quantified by means of the Taylor parameter, combining the major axis of deformation $L$ and the minor axis of deformation $B$
\begin{eqnarray}
D = \frac{L-B}{L+B}.
\label{NS47}
\end{eqnarray}

Here, we study three configurations: (i) a Newtonian droplet in a Newtonian fluid (NN) (ii) a Newtonian droplet in a viscoelastic fluid (NV) and (iii) a viscoelastic droplet in a Newtonian fluid (VN). Fig. \ref{fig:fig3} shows the results for $Re=0.3$, $Ca=0.6$, $Wi=0.4$, $\beta=0.5$, $\lambda_{\mu}=1$, $\lambda_{\rho}=1$ ($Wi=0$ for the NN case).
The domain is discretized with a grid of  $N_{x}\times N_{y}=720\times360$ and $N_{x}\times N_{y}=1440\times720$ corresponding to $Cn=0.038$ and $Cn=0.019$. The Peclet number is set according to $Pe=\frac{3}{Cn}$ to appraoch the sharp-interface limit \citep{Magaletti2013}. 
The results of our simulations  are in good agreement with those in previous studies \citep{Chinyoka2005,Figueiredo2016}, 
where the same problem was solved with a sharp-interface approach (i.e.\ the VOF method). Note, in addition, that the results are not sensitive to $Cn$, see Fig.\ref{fig:fig3}.

\begin{table}[tbp]
\begin{center}
		
\begin{tabular}{c c c c}
\hline
Cn number & RSR & CSR   & DR \\
\hline
0.038  & $720\times360$  & $360\times180$ & $360\times180$  \\
0.019 & $1440\times720$  & $720\times360$ & $720\times360$  \\
\hline
\end{tabular}
\end{center}
\caption{Numerical configurations for the reference single-resolution (RSR), coarse single-resolution (CSR) and dual-resolution (DR) simulations of a droplet deformation in Couette flow. For the DR case, the resolution of the coarser grid is given. The grid size is given as $N_{x}\times N_{y}$ , where $N_i$ indicates the number of grid points in direction $i$.}
\label{Tabel 1}
\end{table}
\begin{figure}[tbp]
	\subfloat[]{\includegraphics[width = 2.8in]{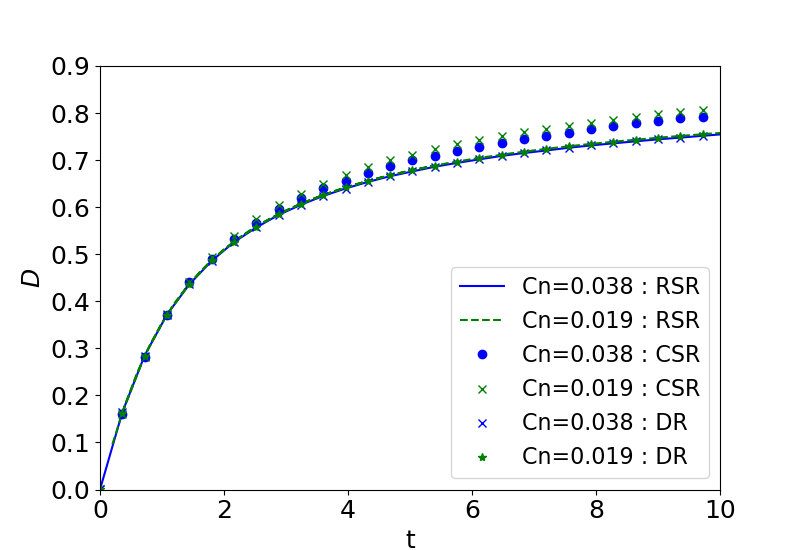}} 
	\subfloat[]{\includegraphics[width = 2.8in]{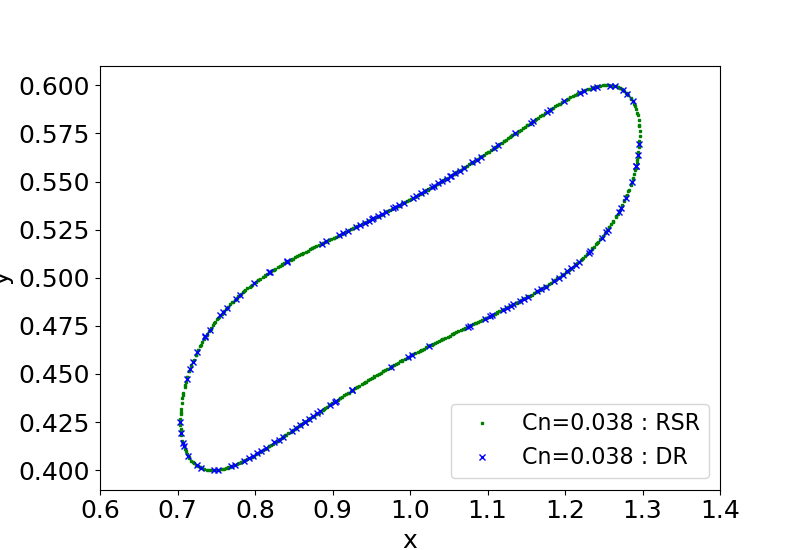}}\\
	\centering
	\hspace*{1.4cm}\subfloat[]{\includegraphics[width = 2.7in]{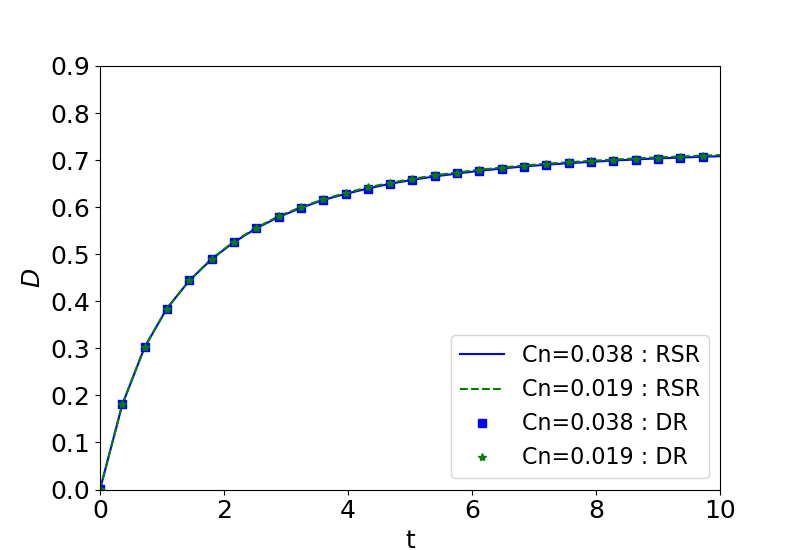}}
	\caption{Comparison between single-resolution and dual-resolution results: (a) Evolution of the droplet deformation parameter $D$ for the $NN$ case. (b) Shape of the droplet at $t=10 s$ for the $NN$ case. (c) Evolution of the droplet deformation parameter $D$ for the $VN$ case.}
	\label{fig:figdual1}
\end{figure}	

\subsubsection{Results from the dual-resolution approach}
We first investigate the convergence and mass conservation of the dual-resolution method for cases NN and NV as introduced in the previous section
using different resolutions; in particular, we use half the resolution used with the single-resolution approach. 
A summary of the simulation parameters is  given in Table \ref{Tabel 1}; it should be recalled that the phase-field indicator function is solved on a twice as fine mesh in the case of dual resolution. To better appreciate the improvements from the dual-resolution strategy, we
also perform simulations with single resolution on the coarser mesh.\\
Fig. \ref{fig:figdual1}(a) 
presents the time evolution of the Taylor deformation parameter for case NN. 
The result obtained with the dual resolution overlaps with the reference (fine) single-resolution data (RSR); 
however, the results obtained with the coarse single-resolution approach (CSR) are significantly different from the reference single-resolution data and the droplet deformation is overpredicted. The shape of the droplets (defined by the contour $\phi=0$)  extracted at $t=10 s$ from the single-resolution and dual-resolution data are compared in Fig.\ref{fig:figdual1}(b): the results shows that the mass (volume) of the droplet is perfectly conserved on the dual-resolution grid.
Finally, we consider case VN, the case of a viscoelastic droplet in a Newtonian fluid. The results pertaining to the Taylor deformation parameter, see Fig.\ \ref{fig:figdual1}(c), are  identical. The results therefore show that the dual resolution approach is able to capture the dynamics of the problem equally well as solving all the variables on the finer mesh; reducing the resolution for the velocities, pressure, and polymer stresses by a factor 2 does not affect the results as long as the phase-field parameter is resolved on a finer mesh.

\begin{figure}[tbp]
	\centering
	\includegraphics[width=0.4\textwidth]{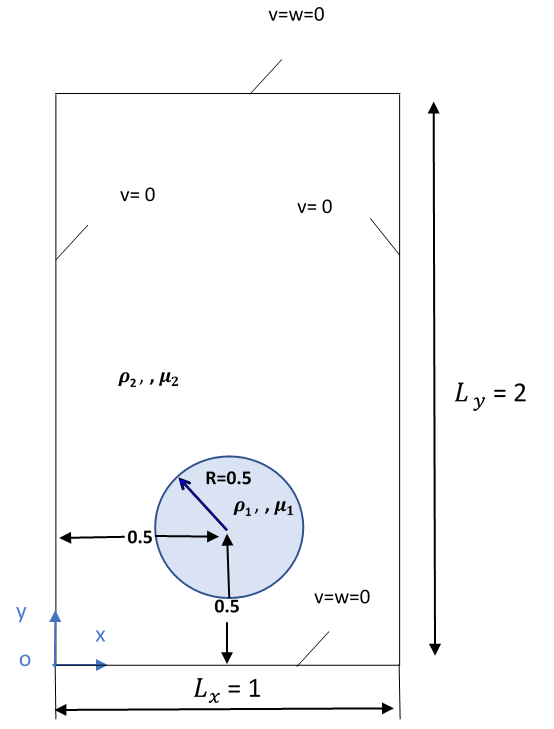}
	\caption{Initial configuration and boundary conditions for the rising bubble benchmark case, see also \cite{Hysing2009}.}
	\label{fig:fig4}
\end{figure}

\begin{table}[tbp]
	\begin{center}
		\caption{\footnotesize{Fluid properties and dimensionless numbers for the rising bubble benchmark.}}
		\label{Tabel2}
		\begin{tabular}{c c c c c c c c c c}
			\hline
			Test case & $\rho_{1}$ & $\rho_{2}$ & $\mu_{1}$ & $\mu_{2}$ & $\sigma$ & $Re$ & $Eo$ & $\lambda_{\mu}$ & $\lambda_{\rho}$\\
			\hline
			1 & 1000  & 100 & 10  & 1 & 24.5  & 35 & 10  & 10 & 10\\
			2 & 1000  & 1 & 10  & 0.1 & 1.96  & 35 & 125  & 100 & 1000\\
			\hline
		\end{tabular}
	\end{center}
\end{table}
\vspace{0.2cm}
		
\subsection{Rising bubble}
\subsubsection{Single-resolution numerical setup}
Next, we consider the rising bubble benchmark problem described in \cite{Hysing2009}. 
The computational domain is a 2D rectangle,  $\Omega=[0,L_{x}]\times[0,L_{y}]$, with the bubble initially placed at $(x,y)=(0.5,0.5)$ and  no-slip boundary condition imposed at the bottom and upper walls and free-slip condition on the side walls, see Fig.\ \ref{fig:fig4}. 
Both the bubble and the surrounding fluids are Newtonian with viscosity ratio $\lambda_{\mu}=\frac{\mu_{2}}{\mu_{1}}$ density ratio $\lambda_{\rho}=\frac{\rho_{2}}{\rho_{1}}$. The bubble rises in the y-direction due to a smaller density $(\rho_{1}<\rho_{2})$ than the ambient fluid under the gravitational force $(g_{x},g_{y})=(0,-0.98)$.

The Reynolds number $Re=\frac{\rho_{2}u_{g}2R}{\mu_{2}}$, E{\"o}tv{\"o}s number (Eo) number $Eo=\frac{2\rho_{2}u_{g}^{2}R}{\sigma}$, the Cahn number, $Cn=\frac{\eta}{2R}$, and the Peclet number $Pe=\frac{2\sqrt{2}u_{g}2R\eta}{3M\sigma}$ are defined using $L_{ref}=2R$ and $u_{g}=\sqrt{2gR}$ as the length and velocity scales. The Peclet number is obtained according to the scaling $Pe=\frac{3}{Cn}$ \citep{Magaletti2013}. The fluid properties and the corresponding dimensionless numbers used in the simulation are given in Table \ref{Tabel2}.

To quantify the results, we follow the position of the bubble center of mass:
\[ y_{c} = \frac{\int\limits_{\Omega} y \ d\Omega}{\int\limits_{\Omega}\ d\Omega},  \]
\vspace{0.2cm}
where y is the vertical coordinate, and the bubble rising velocity:
\[ v_{c} = \frac{\int\limits_{\Omega} v_{y} \ d\Omega}{\int\limits_{\Omega}\ d\Omega}  \]
where $v_{y}$ is the vertical component of the velocity field $\mathbf{u}$.

\begin{figure}[tbp]
	\centering
	\hspace*{1.3cm}\subfloat[]{\includegraphics[width = 2.6in]{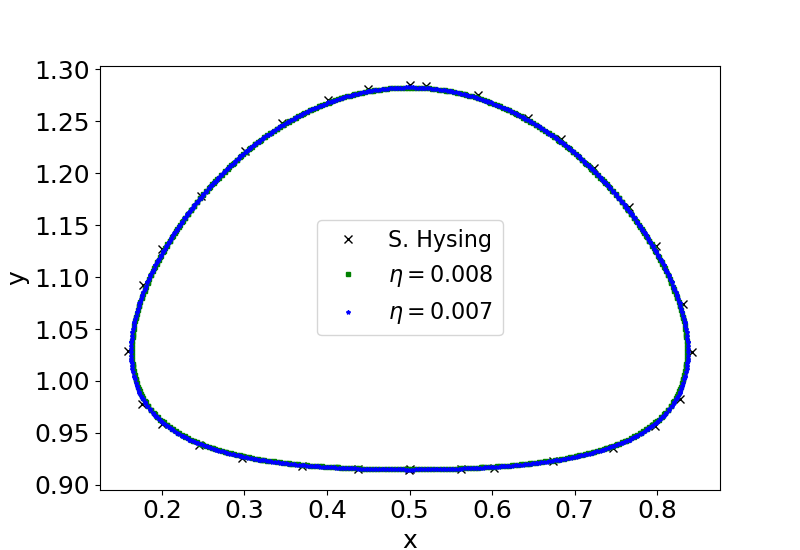}}\\
	\subfloat[]{\includegraphics[width = 2.6in]{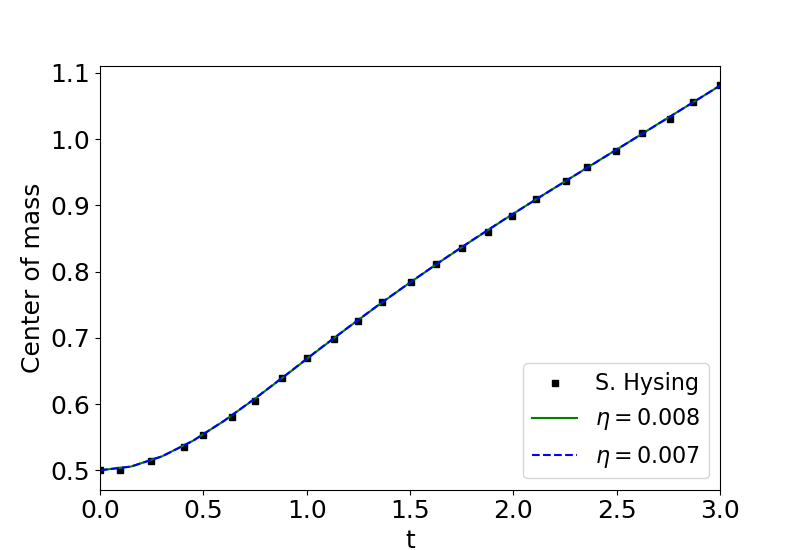}} 
	\subfloat[]{\includegraphics[width = 2.6in]{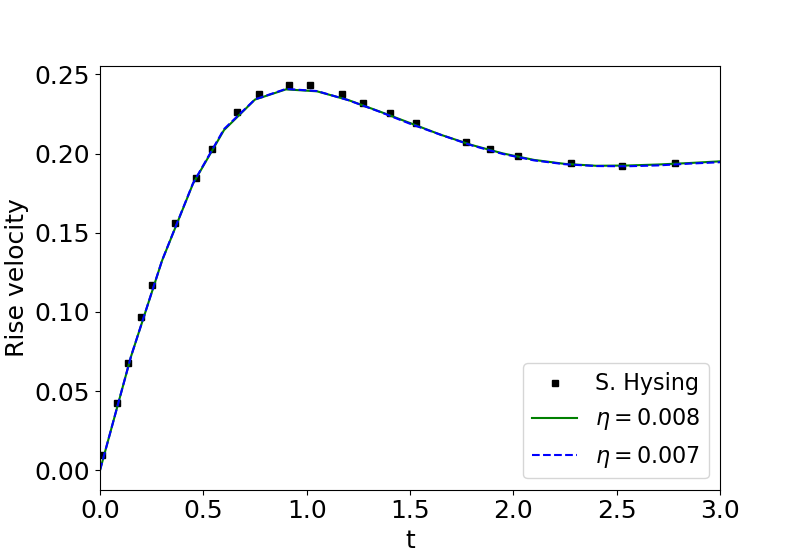}}
	\caption{Rising bubble, test case 1: (a)  Final shape of the bubble (b) Time trace of the center of mass location; (c) Bubble rising velocity. Our results are compared with the results in \cite{Hysing2009}.}
	\label{fig:fig5}
\end{figure}

\subsubsection{Single resolution results: Test case 1}
In this case, the bubble rises and stretches horizontally with limited deformation due to the high surface tension, see the parameter setting in Table \ref{Tabel2}. We use two values of the interface  width, $\eta=0.008$ and $\eta=0.007$, and compare the bubble observables against the reference \citep{Hysing2009} in Fig.\ \ref{fig:fig5}. 
Panel (a) displays the bubble  shape, panel (b) the time trace of the centre of mass location and (c) the bubble rising velocity.
For all quantities,  an excellent agreement with literature data is seen.

\begin{figure}[tbp]
	\centering
	\hspace*{1.3cm}\subfloat[]{\includegraphics[width = 2.6in]{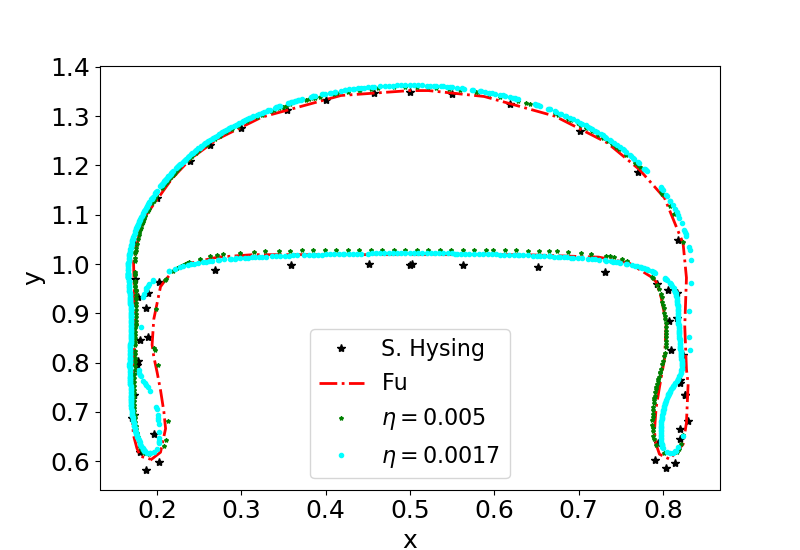}}\\
	\subfloat[]{\includegraphics[width = 2.6in]{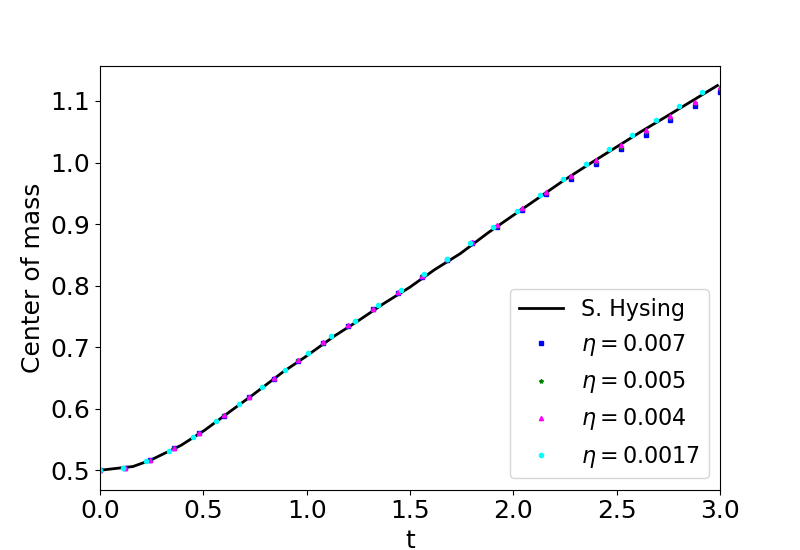}} 
	\subfloat[]{\includegraphics[width = 2.6in]{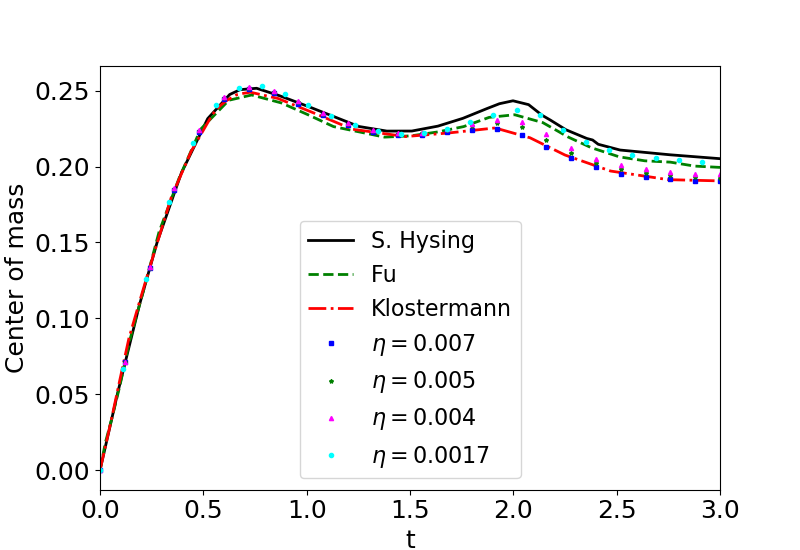}}
	\caption{Rising bubble, test case 2: (a)  Final shape of the bubble (b) Time trace of the center of mass location; (c) Bubble rising velocity. Our results are compared with the results in \cite{Hysing2009,Klostermann2013,Fu2020}.}
	\label{fig:fig6}
\end{figure}
\subsubsection{Single resolution results: Test case 2}
In test case 2, a lower surface tension causes the bubble to deform more and thin filaments to appear as the bubble rises. 
In this case, the different observables are compared 
against the results of three previous studies \citep{Hysing2009,Klostermann2013,Fu2020} in Fig.\ \ref{fig:fig6}. 
Four interface widths are compared for the results presented here.
In particular, we compare the bubble final shape at $t=3s$ with \cite{Hysing2009} (FreeLIFE solver, using the level set approach) and \cite{Fu2020} (Phase-filed method) in Fig. \ref{fig:fig6}(a). 
The agreement is excellent with \cite{Fu2020} when using the same interface  width ($\eta=0.005$), and the shape of the bubble converges towards the Level-Set method results by \cite{Hysing2009},
with the filaments becoming thinner when decreasing the interface width to $\eta=0.0017$. 
Fig. \ref{fig:fig6}(b) presents the comparison of the center of mass position, and again the results are in good agreement with \cite{Hysing2009} when decreasing the interface width to $\eta=0.0017$. All the methods predict almost the same rising velocity up to $t=1.5s$, while each reference reports slightly different rising velocity after $t=1.5s$ depending on the method used. Our result predicts the same rising velocity as \cite{Fu2020} when choosing the same interface width and converges to the velocity in \cite{Klostermann2013} (VOF) and \cite{Hysing2009}  when using a smaller interface  width as shown in Fig. \ref{fig:fig6}(c).

\begin{figure}[tbp]
	\centering
	\hspace*{1.3cm}\subfloat[]{\includegraphics[width = 2.6in]{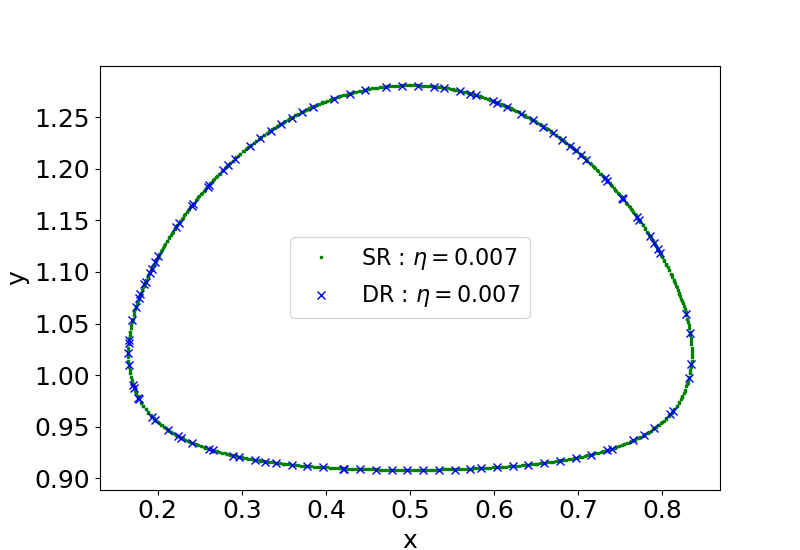}}\\
	\subfloat[]{\includegraphics[width = 2.6in]{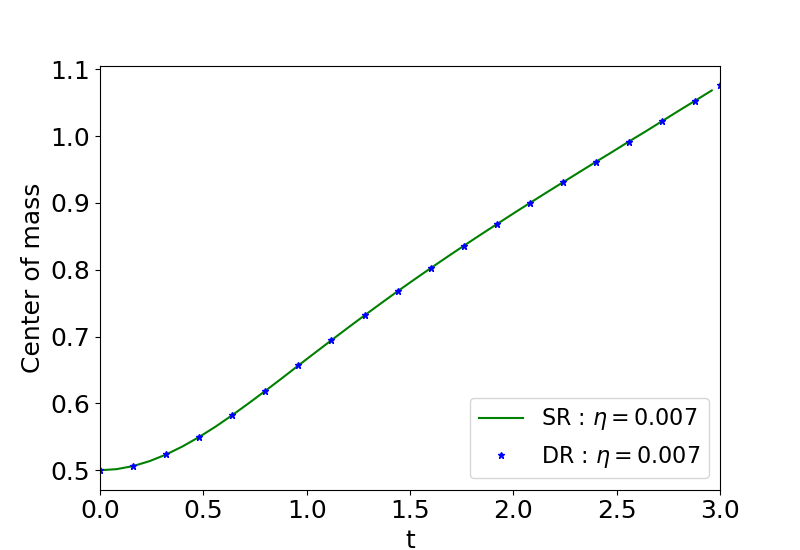}} 
	\subfloat[]{\includegraphics[width = 2.6in]{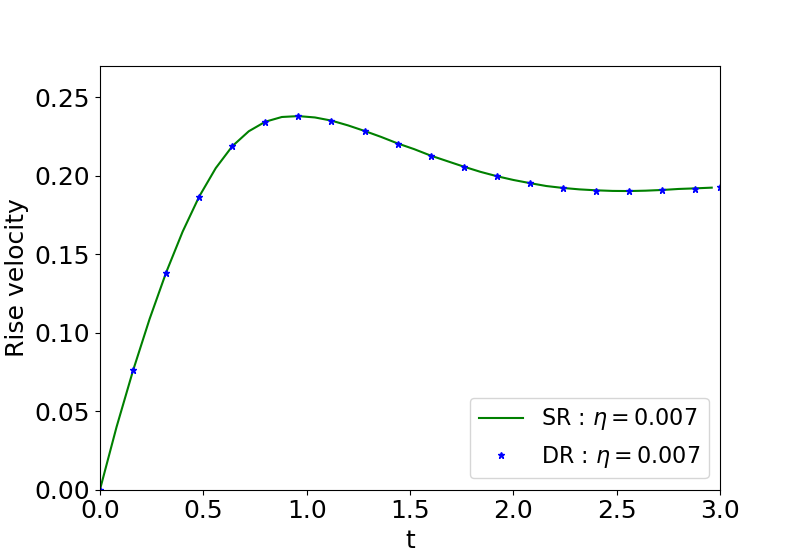}}
	\caption{Comparison between single- and dual-resolution results for the rising bubble case 1: (a) final shape of the bubble (b) time history of the center of mass vertical position and (c) bubble rising velocity.}
	\label{fig:figdual2}
\end{figure}

\subsubsection{Dual-resolution results}

By the rising bubble test case, we test the performance of the dual-resolution approach in the presence of the density and viscosity contrasts. All the physical parameters are the same as for the single-resolution results just discussed, except for the grid-resolution.  Table \ref{Tabel 3} reports the resolutions adopted with the two methods.
\begin{table}[tbp]
\begin{center}
\begin{tabular}{c c c c c}
\hline
Case   &  capillary width $\eta$ & SR & DR \\
\hline
Case 1 & 0.007 & $320\times640$  & $160\times320$ \\
Case 2 & 0.0017 & $1200\times2400$ & $600\times1200$  \\
\hline
\end{tabular}
\end{center}
\caption{Computational configuration for the single-resolution (SR) and dual-resolution(DR) simulations pertaining to the rising bubble test case from \cite{Hysing2009}. The grid size is given as $N_{x}\times N_{y}$.}
\label{Tabel 3}
\end{table}

We start from test case 1. Fig. \ref{fig:figdual2} (a) presents the bubble shape at $t=3s$ (defined by the $\phi=0$ contour of the phase-field parameter) as obtained on dual-resolution and single-resolution grids, where one can see that the shapes are almost the same. 
The center of mass and the rising velocity are displayed in Fig. \ref{fig:figdual2} (b) and (c) respectively, again showing negligible differences between the two approaches.

\begin{figure}[!htb]
\centering
\hspace*{1.3cm}\subfloat[]{\includegraphics[width = 2.6in]{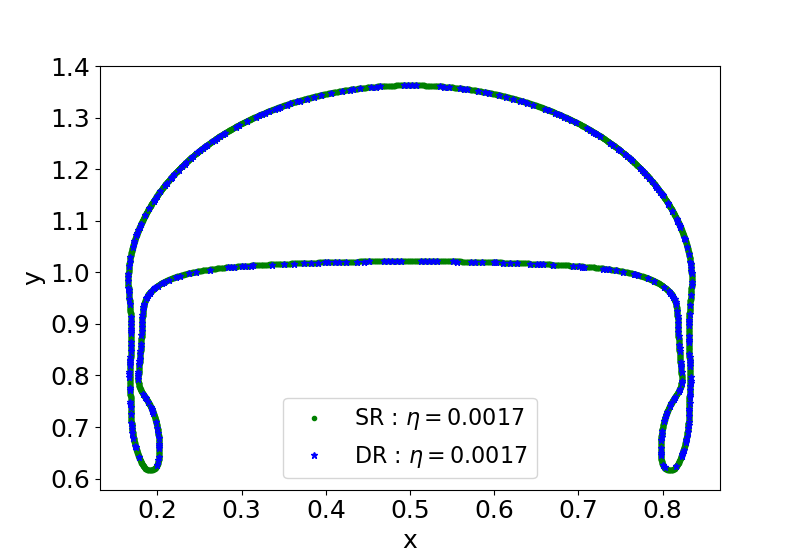}}\\
\subfloat[]{\includegraphics[width = 2.6in]{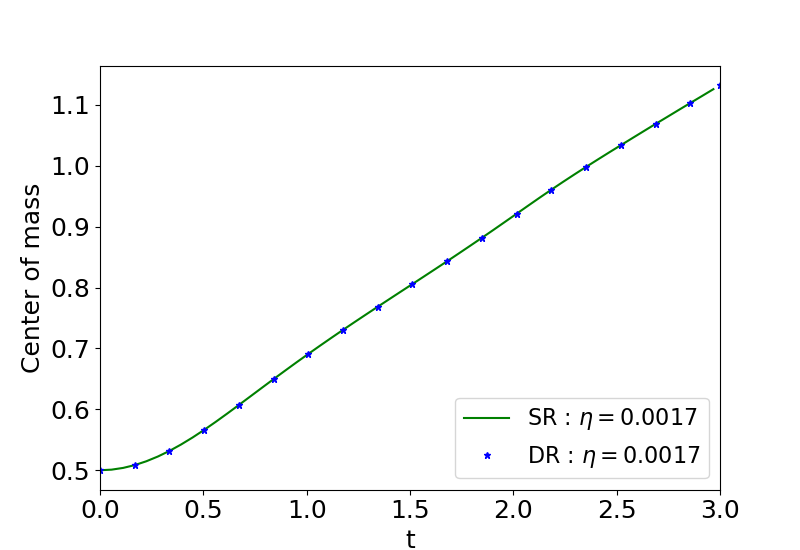}} 
\subfloat[]{\includegraphics[width = 2.6in]{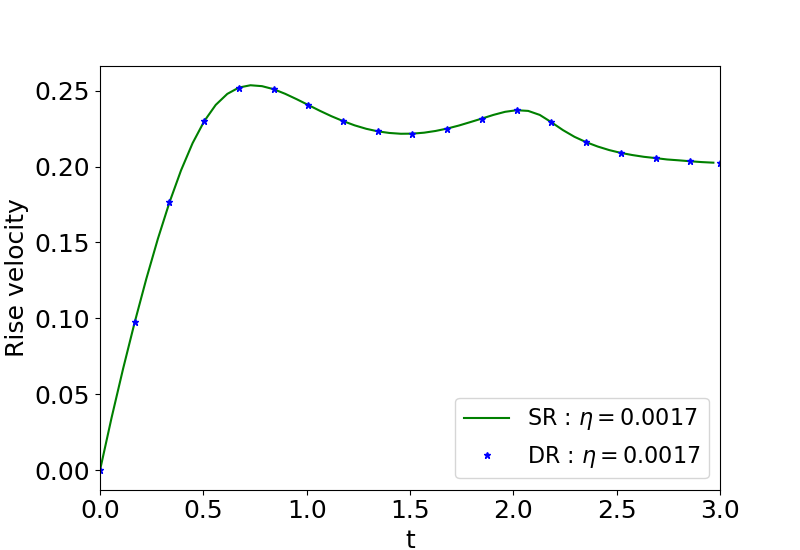}}
\caption{Comparison between single- and dual-resolution results for  the rising bubble case 2: (a) final shape of the bubble (b) time history of the center of mass vertical position and (c) bubble rising velocity.}
\label{fig:figdual3}
\end{figure}

Regarding test case 2, presenting large droplet deformations, we compare results for the interface thickness $\eta=0.0017$. 
Fig.\ \ref{fig:figdual3} (a) presents the comparison of the bubble final shape at $t=3s$, whereas Fig. \ref{fig:figdual3} (b) and Fig. \ref{fig:figdual3} (c) display the center of mass vertical position and the bubble rising velocity. 
The results confirm that the dual-resolution approach correctly captures the bubble dynamics also in the case of large density and viscosity ratios.

\begin{figure}[tbp]
\centering
\includegraphics[width=0.8\textwidth]{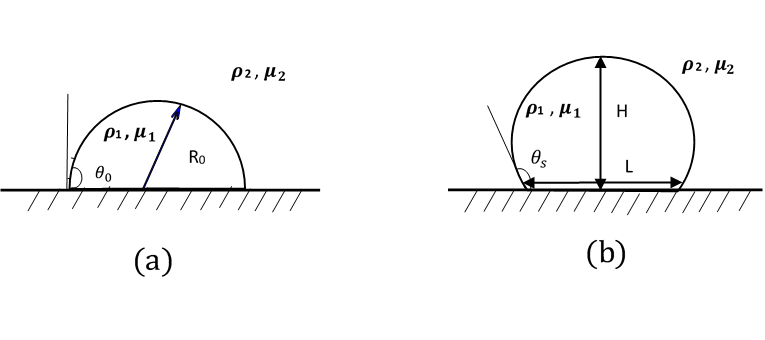}
\caption{ (a) The initial shape of the drop and (b) the equilibrium shape}
\label{fig:fig7}
\end{figure}

\begin{figure}[tbp]
\centering
\includegraphics[width=0.55\textwidth]{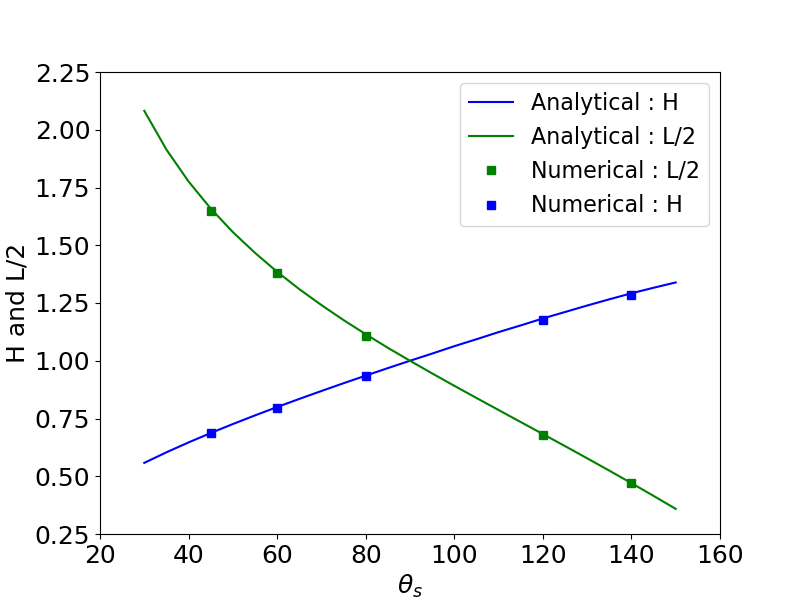}
\caption{Comparison between the analytical \citep[see][]{Dong2012} and numerical values of the spreading length $L$ and height $H$ as a function of the static contact angle $\theta_{s}$. }
\label{fig:fig8}
\end{figure}

\subsection{Wetting of a Newtonian droplet}
\subsubsection{2D equilibrium}
In this subsection, we consider a Newtonian droplet on a horizontal solid surface and  neglect gravity  to validate the static contact-angle, Neumann and periodic boundary conditions ($D_{w}=0$). We use a 2D rectangular domain $\Omega=[0,6]\times[0,1]$ with a semi-circular droplet of radius $R_{0}=0.5$ and initial contact angle $\theta_{0}=90^{0}$ placed at $(x,y)=(3,0)$, see Fig. \ref{fig:fig7}(a). The boundary conditions are no-slip and no-penetration on the two walls and periodic in the x-direction. 
Different static contact angles $\theta_{s}$ are used in the simulation, and the droplet will spread or recoil depending on $\theta_{s}$. Here, fluid $1$ indicates the droplet phase and fluid $2$ the surrounding medium. Using the velocity scale $u_{ref}=\frac{\sigma}{\mu_{1}}$ and the properties of the droplet as reference values we consider the following values of the relevant dimensionless numbers :
\begin{equation}{Re=\frac{\rho_{1}\sigma R_{0}}{\mu_{1}^{2}}=10, \quad Ca=1, \quad Cn=\frac{\eta}{R_{0}}=0.02, \quad \lambda_{\mu}=\frac{\mu_1}{\mu_2}=1.1, \quad \lambda_{\rho}=\frac{\rho_1}{\rho_2}=1.1}\end{equation}
The mobility parameter is chosen by the criterion $Cn\leq4S$ to attain the sharp-interface limit in the computation \citep{Yue2010}, where $S=\frac{\sqrt{M\sqrt{\mu_{1}\mu_{2}}}}{R}$ is the ratio between the diffusion length $l_{d}=\sqrt{M\sqrt{\mu_{1}\mu_{2}}}$ and the drop lenght scale.\\
When the droplet reaches the equilibrium condition, the height of the drop $H$ and the spreading length $L$ can be derived by mass (volume) conservation  \citep[see e.g.][]{Dong2012} 

\begin{align}
\label{NS48}
\begin{split}
	H&=R_{0}[1-\cos{\theta_{s}}]\sqrt{\frac{\pi}{2(\theta_{s}-\sin{\theta_{s}}\cos{\theta_{s}})}} 
	\\
	L&=2R_{0}\sin{\theta_{s}}\sqrt{\frac{\pi}{2(\theta_{s}-\sin{\theta_{s}}\cos{\theta_{s}})}}.
\end{split}
\end{align}

Fig. \ref{fig:fig8} shows the comparison between the analytical and numerical values of $L$ and $H$ for different values of the static contact angle $\theta_{s}$, confirming the accuracy of the present numerical approach.

\captionsetup[subfigure]{labelformat=empty}
\begin{figure}[tbp]{}
	\subfloat[$t^{*}=0$]{\includegraphics[width = 2.5in]{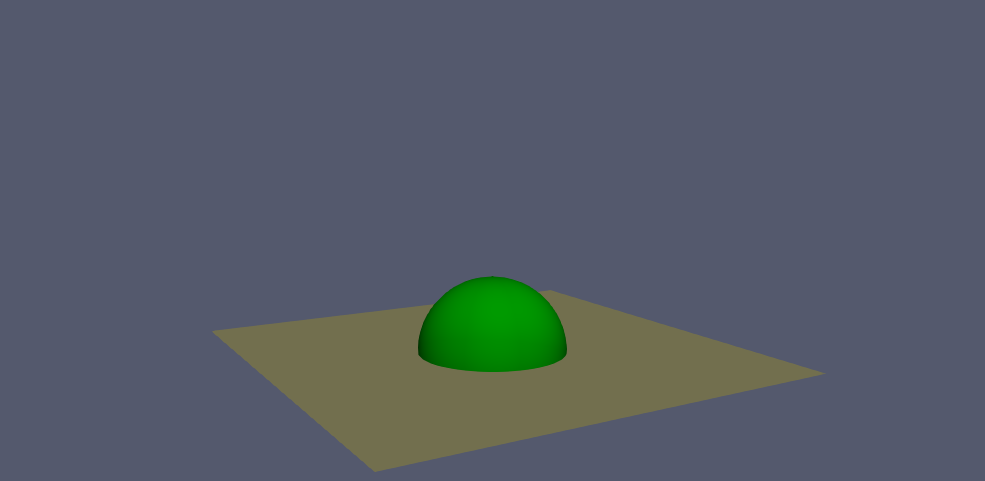}}
	\centering
	\hspace*{0.5cm}
	\subfloat[$t^{*}\approx54$]{\includegraphics[width = 2.5in]{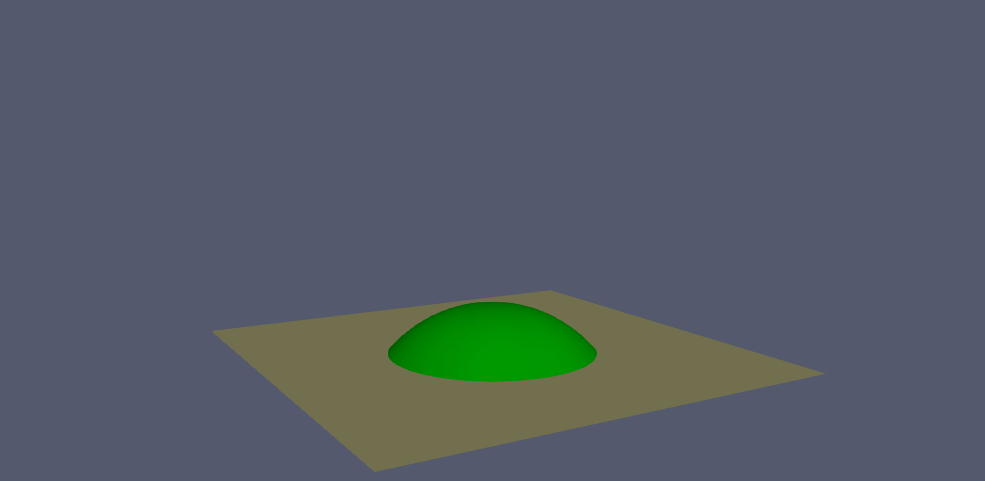}}\\
	\centering
	\subfloat[$t^{*}\approx107$]{\includegraphics[width = 2.5in]{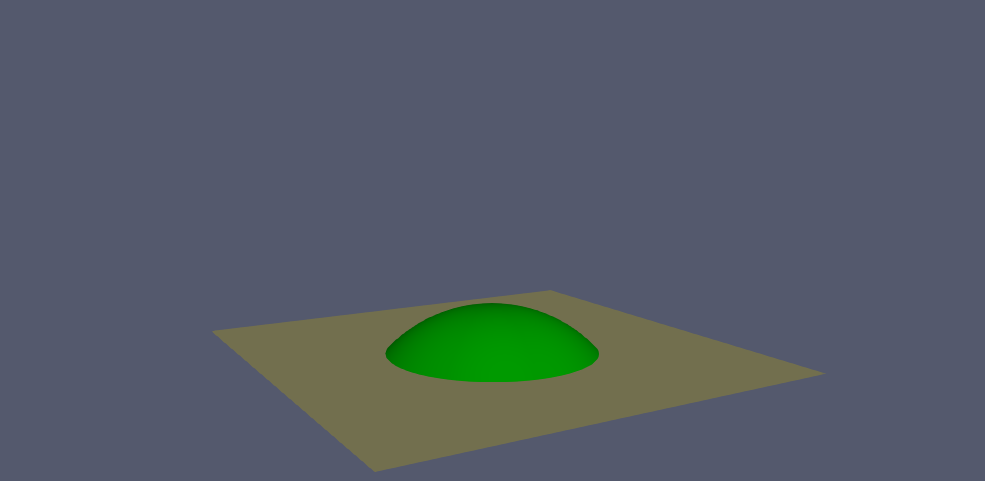}}
	\hspace*{0.5cm}
	\subfloat[$t^{*}\approx172$]{\includegraphics[width = 2.5in]{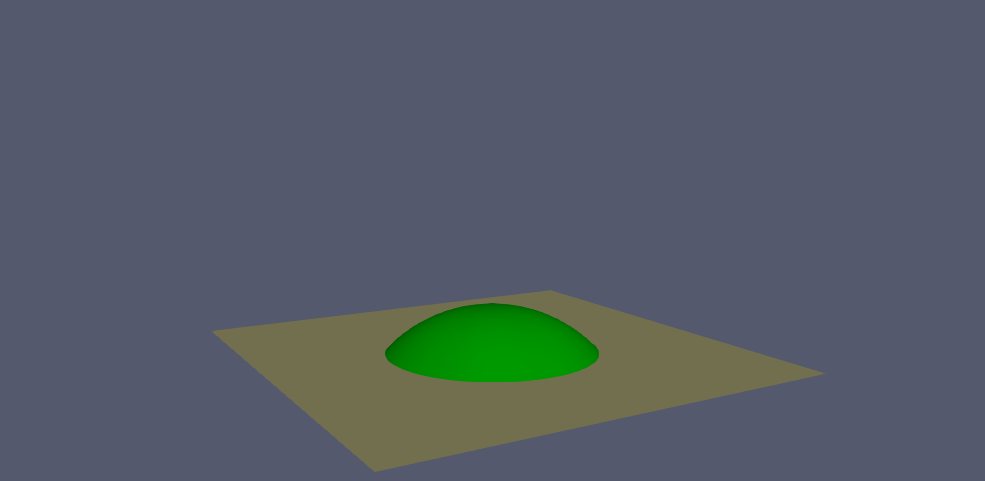}}
	\caption{Spreading of a 3D droplet on a solid substrate with static contact angle $\theta_{s}=45^{o}$.}
	\label{fig:fig9}
\end{figure}

\begin{figure}[tbp]
\centering
\includegraphics[width=0.55\textwidth]{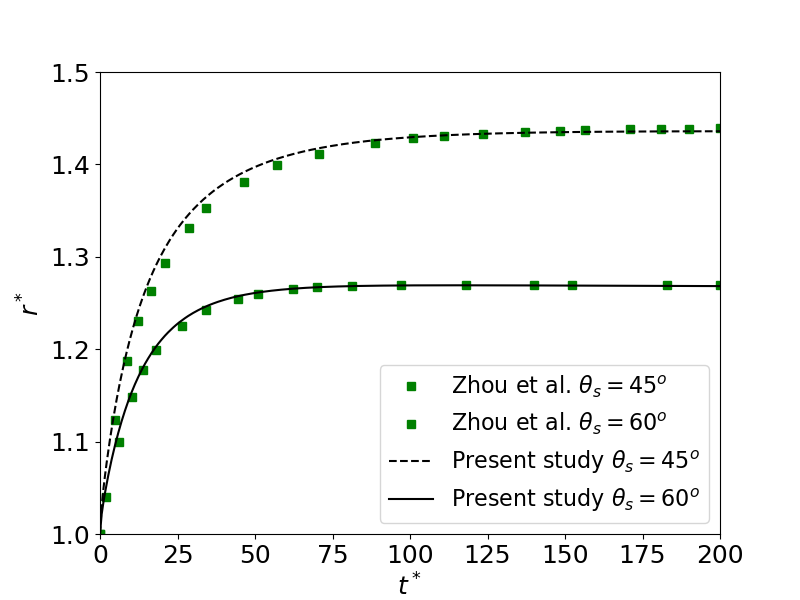}
\caption{Time evolution of the spreading radius of a 3D droplet on a solid substrate with static contact angle $\theta_{s}=45^{o}$ and  $\theta_{s}=60^{o}$.
	Our results are compared against the data in \cite{Zhou2009}.
	The spreading radius and time have been made dimensionless by the droplet initial radius and the capillary time $t_{r}=\frac{\mu_{1}R_{0}}{\sigma}$.}
\label{fig:fig10}
\end{figure}

\subsubsection{3D drop spreading on a hydrophilic substrate}

As a 3D test case, we consider the spreading of a  Newtonian drop on a horizontal solid surface as in \cite{Zhou2009}, with the
static contact-angle boundary condition, i.e.\ $D_{w}=0$ in Eq. \ref{NS15}. 
The computational domain has size $\Omega=[0,6]\times[0,6]\times[0,3]$, with a semi-spherical droplet of radius $R_{0}=1$ at $(x,y,z)=(3,3,0)$ and an initial contact angle $\theta_{0}=90^{0}$. The no-slip and no-penetration conditions are imposed on the bottom and top walls, respectively, and periodic boundary conditions are applied for all variables in the lateral directions. Two static contact angles are considered, $\theta_{s}=60^{o}$ and $\theta_{s}=45^{o}$. 
We choose the following dimensionless numbers, defined with velocity scale $u_{ref}=\frac{\sigma}{\mu_{1}}$ and the fluid properties as reference values:  
\begin{align*}
	Re&=0.05,  &  Ca&=1,  &   Cn&=0.03,  &   Pe&=2828,  &  \lambda_{\mu}&=1,   &  \lambda_{\rho}&=1.
\end{align*}
Visualizations of the droplet spreading for the static contact angle $\theta_{s}=45^{o}$ are presented in Fig.\ \ref{fig:fig9}. 
The droplet with initial contact angle $\theta_{0}=90^{o}$ spreads on the surface to reach the final equilibrium  with contact angle $\theta_{s}=45^{o}$.  

The comparison with the results by \cite{Zhou2009} is reported in Fig. \ref{fig:fig10} in terms of evolution of the droplet radius, showing good agreement between the two numerical solutions.

\begin{figure}[tbp]
	\centering
	\includegraphics[width=0.55\textwidth]{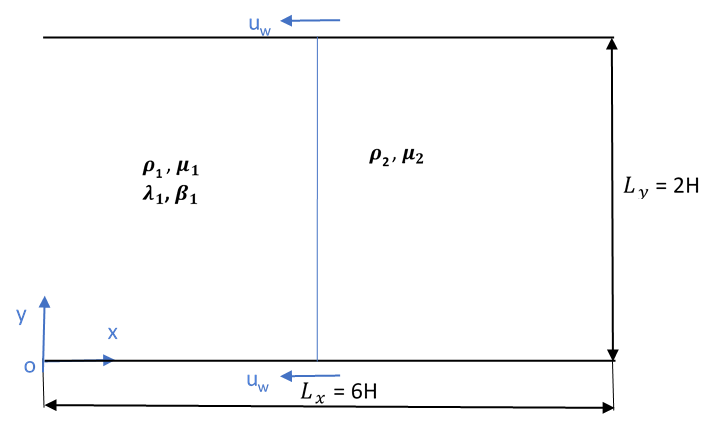}
	\caption{Initial configuration for the displacement of an Oldroyd-B and Newtonian fluid. The interface between the two fluids is initially vertical and placed at $x=3H$, see also text and  \cite{Yue2012} for the definition of this test case.}
	\label{fig:fig11}
\end{figure}

\begin{figure}[tbp]
	\centering
	\subfloat[\textbf(a)]{\includegraphics[width = 2.8in]{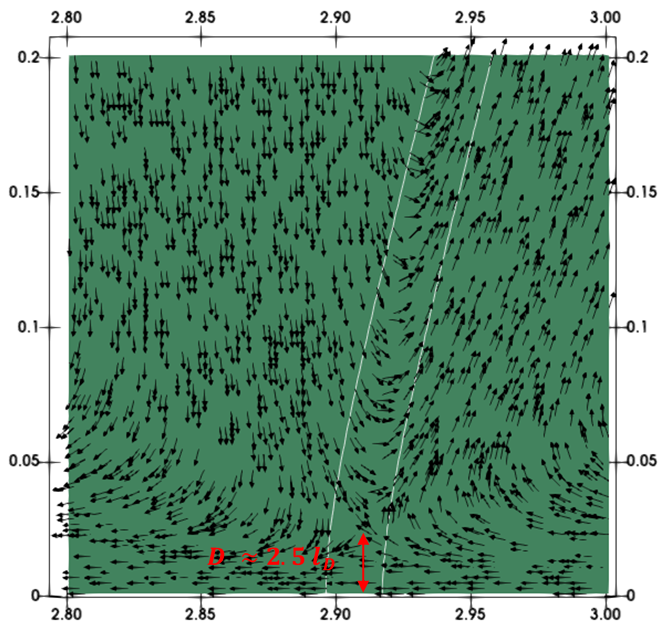}} 
	\hspace*{0.5cm}
	\subfloat[\textbf(b)]{\includegraphics[width = 2.8in]{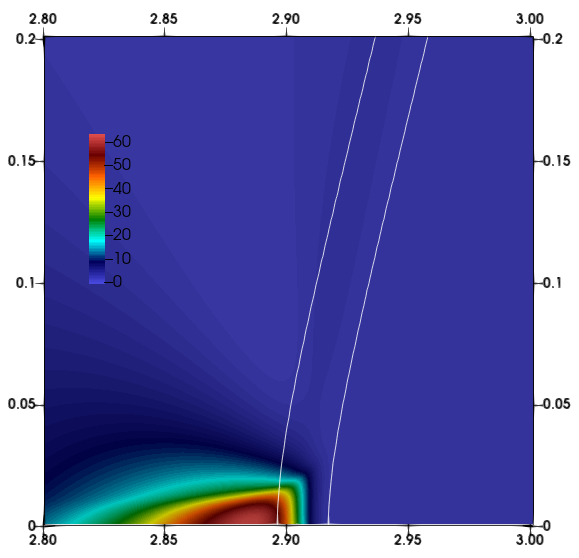}}
	\caption{The flow field around the contact line in the case of the displacement of an Oldroyd-B and Newtonian fluid, with $Ca=0.02$ and $Wi=0.02$. (a) The velocity field around the contact line. The two white curves indicate the contour $\phi=\pm 0.9$, and the stagnation point location is $D=2.5 l_d$ as indicated. (b) Contours of the dimensionless polymer stress $\tau_{xx}^{*}$ from the present study. Both results are in good agreement with \cite{Yue2012}.}
	\label{fig:fig12}
\end{figure}

\begin{figure}[tbp]
	\centering
	\includegraphics[width=0.6\textwidth]{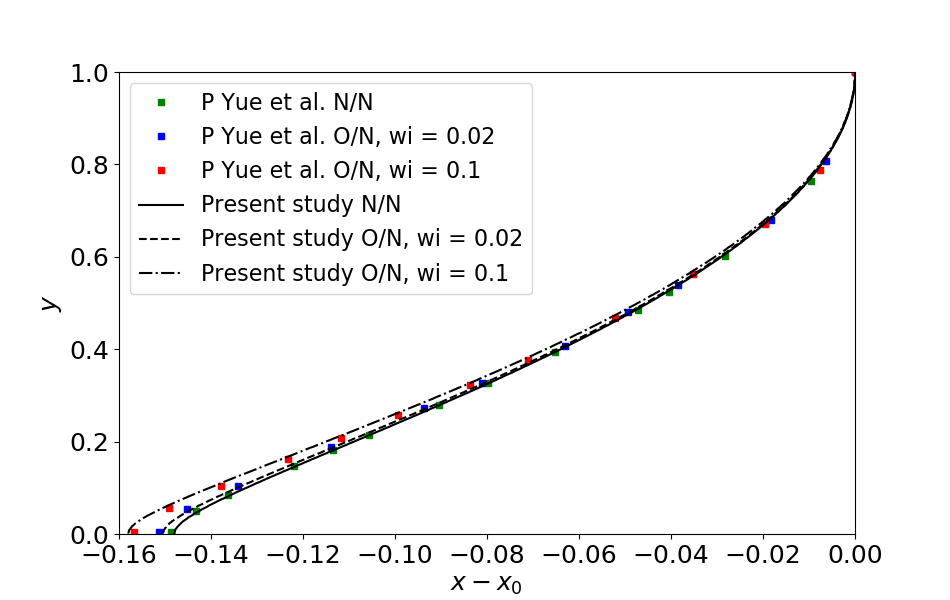}
	\caption{Comparison of interface shape when an Oldroyd-B fluid displaces a Newtonian fluid in a 2D channel. The current results are compared with the data in \cite{Yue2012}. }
	\label{fig:fig13}
\end{figure}

\subsection{Wetting of a viscoelastic fluid}

Finally, we validate the numerical results for the wetting of viscoelastic fluids against the benchmark in \cite{Yue2012}. 
The computational domain is 2D, of size $\Omega=[0,6H]\times[0,2H]$. The  interface  between  two  phases  is  initially  vertical  and placed at the center of the channel ($x=3H$). The advancing fluid on the left side is an Oldroyd-B fluid, whereas the receding fluid is a Newtonian fluid, configuration denoted $O/N$ in \cite{Yue2012}, see Fig. \ref{fig:fig11}. 
At $t=0$, we impose a parabolic velocity profile at the inlet and outlet, with the two walls moving with velocity $u_{w}=u_{av}$ to the left. In other words, the wall velocity $u_{w}$ is chosen so that the net flow rate is zero and the interface has a stationary profile at the steady-state condition.
The no-slip and no-penetration boundary conditions are imposed on the two walls, with static contact angle ($D_{w}=0$) $\theta_{s}=90^{0}$.

The dimensionless numbers defining this problem are the Reynolds number  $Re=\rho_{1}u_{w}H/\mu_{1}=0.01$, the Capillary number $Ca=\mu_{1}u_{w}/{\sigma}=0.02$, $Wi=\lambda_{H}u_{w}/{H}=0.02$, the dimensionless diffusion length  $S=\sqrt{M\sqrt{\mu_{1}\mu_{2}}}/{R}=0.01$; the Cahn number $Cn=\eta/R=0.05$, the relaxation time ratio $\beta=\mu_{s}/\left(\mu_{s}+\mu_{p}\right)=0.5$, the viscosity ratio $\lambda_{\mu}=\mu_{2}/{\mu_{1}}=1$ and the density ratio $\lambda_{\rho}=\rho_{2}/{\rho_{1}}=1$.

Fig \ref{fig:fig12} presents the contact line position and the polymer stress at steady state for $Wi=0.02$. In this test case, the no-slip boundary condition is imposed on the wall, and the motion of the contact line is entirely due to the Cahn–Hilliard diffusion. The diffusion length $l_{d}=\sqrt{M\sqrt{\mu_{1}\mu_{2}}}$ can be related to the slip length $l_{s}$ in the sharp-interface limit \citep{Yue2010}, and the stagnation point location $D\approx2.5l_{D}$ in Fig.\ref{fig:fig12}(a) is in good agreement with their result. Panel (b) shows the dimensionless polymer stress $\tau_{xx}^{*}=\left(\tau_{xx}H\right)/\left(\mu_{1}u_{w}\right)$ around the contact line, also in good agreement. 

Note that the elasticity of the fluid enhances the wetting when the advancing fluid is a viscoelastic fluid \citep{Yue2012,Wang2015}. Fig. \ref{fig:fig13} shows that indeed the elasticity of the Oldroyd-B fluid causes the interface to bend more into the displacing fluid, in agreement with \cite{Yue2012}. 

\begin{figure}[tbp]
	\centering
	\includegraphics[width=0.6\textwidth]{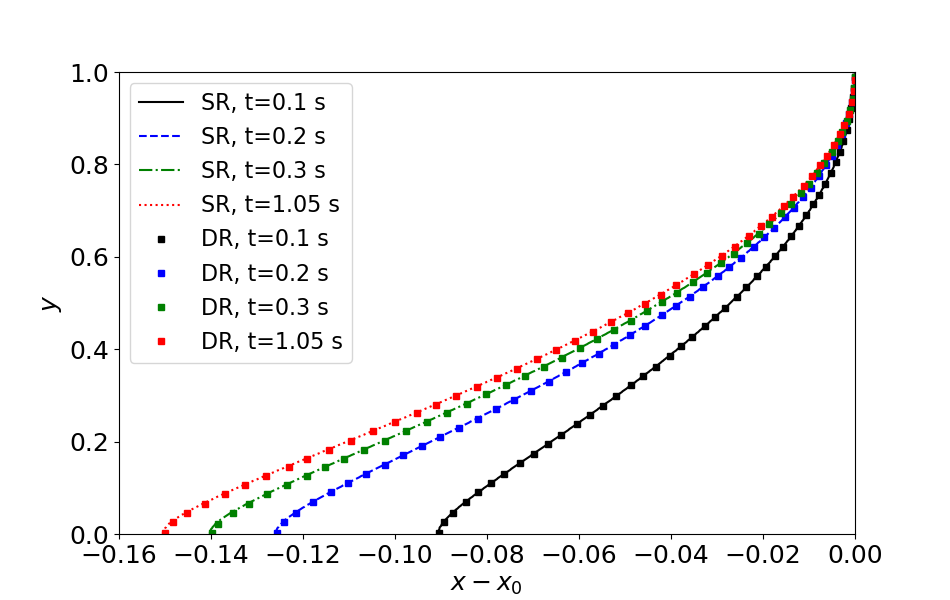}
	\caption{Time evolution of the interface shape as obtained with the single-resolution and dual-resolution approach for the case of  an Oldroyd-B fluid displacing a Newtonian fluid, with $Re=0.01$, $Wi=0.02$, and $Ca=0.02$.}
	\label{fig:figdual4}
\end{figure}

\subsubsection{Wetting of a viscoelastic fluid: dual-resolution results }

Next, we examine the performance of the dual-resolution approach in viscoelastic fluid wetting. 
We use the same parameters used in the simulation with single resolution except for the mesh size.

The number of grid points is $N_{x}\times N_{y}=3000\times1000$ for the single-resolution simulation and $N_{x}\times N_{y}=1500\times500$ for the dual-resolution run (note that the phase-field variable is solved on a twice finer grid). 

The time evolution of the interface location is displayed in Fig. \ref{fig:figdual4} for both resolutions: here, one can observe an excellent agreement between the results obtained with the dual-resolution and single-resolution approach. 

\begin{figure}[tbp]
	\centering
	\includegraphics[width=0.55\textwidth]{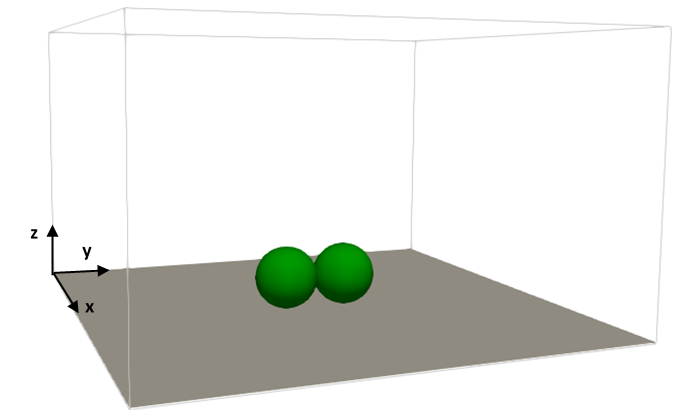}
	\caption{Sketch of the computational domain for the coalescence of the two drops on a superhydrophobic surface}
	\label{fig:fig14}
\end{figure}

\begin{figure}[tbp]
	\centering
	\subfloat[\textbf(a)]{\includegraphics[width = 2.8in]{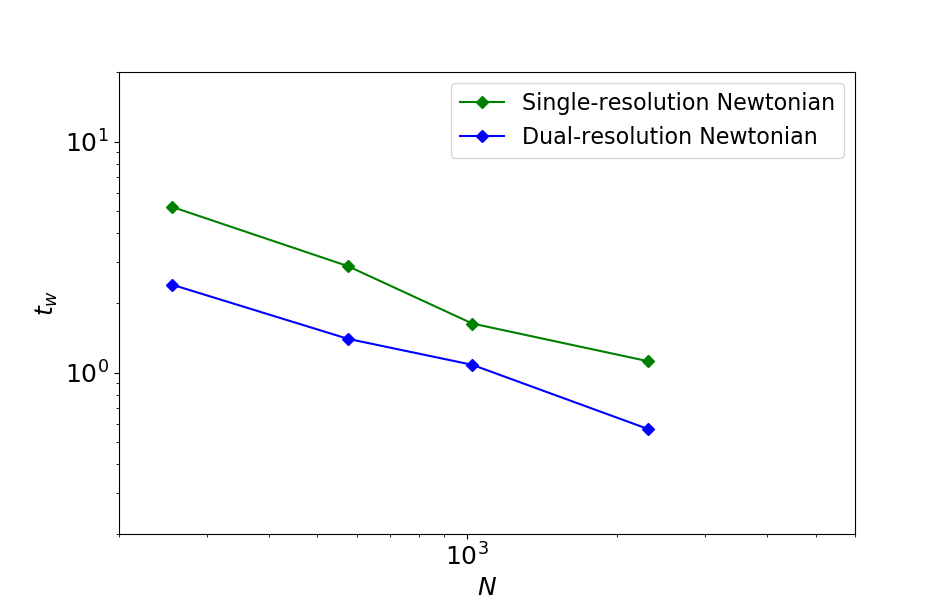}} 
	\hspace*{0.5cm}
	\subfloat[\textbf(b)]{\includegraphics[width = 2.8in]{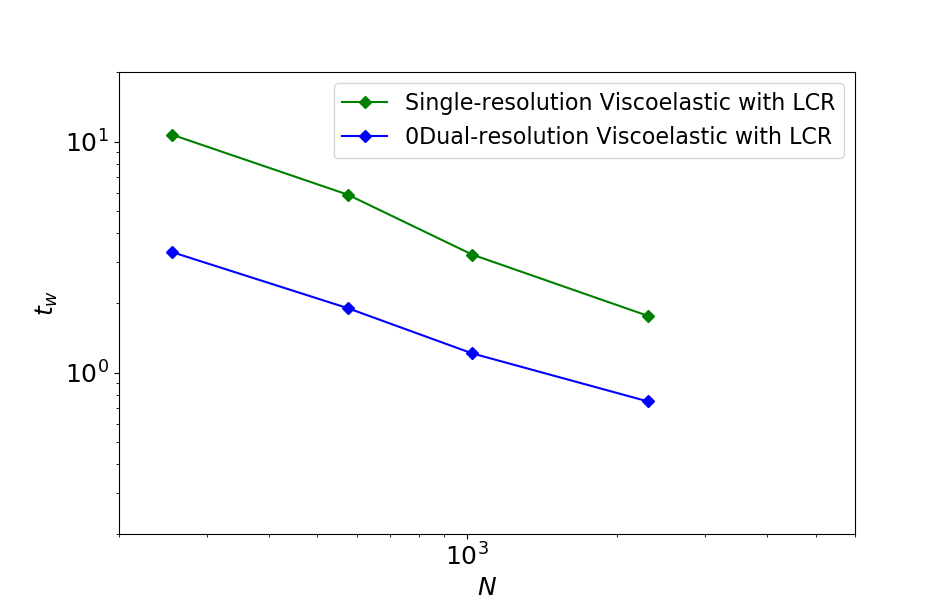}}
	\caption{Strong scaling of the dual-resolution and single-resolution computations for (a) two coalescing Newtonian drops and (b) two viscoelatic drops.  
		$t_{w}$ denotes the wall-clock time in seconds and $N$ is the number of cores ranging from 256 to 2304.}
	\label{fig:figdual5}
\end{figure}
\subsection{Computational performance}

We test the scaling of our implementation by comparing the wall-clock time per iteration for the dual-resolution and single-resolution simulations. We consider the coalescence of  two Newtonian and viscoelastic droplets on a superhydrophobic surface. The computational domain is a 3D box of size $\Omega=[0,6r_{0}]\times[0,6r_{0}]\times[0,6r_{0}]$ discretized with $N_{x}\times N_{y}\times N_{z}=768\times768\times768$ and $N_{x}\times N_{y}\times N_{z}=384\times384\times384$ grid points for the single-resolution and dual-resolution runs. 
The boundary conditions are periodic in the two horizontal directions ($x$ and $y$) and walls in the third direction ($z$). We impose no-slip and no-penetration at the two walls with static contact angle $\theta_{s}=180^{o}$ at the superhydrophobic bottom wall, and $\theta_{s}=90^{o}$ at the top wall, corresponding to a Neumann condition. 
Two initially tangential spherical drops with radius $r_{0}$ are placed in the middle of the domain at the bottom wall $z=0$, and start to coalesce due to diffusion, see Fig. \ref{fig:fig14}. The problem is defined by the Ohnesorge number, defined as $Oh=\frac{\mu_{d}}{\sqrt{\rho_{d}\sigma r_{0}}}$ choosing capillary-inertia velocity as the velocity scale $u_{ci}=\sqrt{\frac{\sigma}{\rho_{d}r_{0}}}$. 
The dimensionless numbers in the simulation are:
\begin{align*}
	Oh&=\frac{\mu_{d}}{\sqrt{\rho_{d}\sigma r_{0}}}=0.118,  &  Wi&=\frac{\lambda_{H}u_{ci}}{r_{0}}=10,  &   Cn&=\frac{\eta}{r_{0}}=0.015  \\   Pe&=\frac{2\sqrt{2}u_{ci}r_{0}\eta}{3M\sigma}=\frac{6}{Cn}, &
	\beta&=\frac{\mu_{s}}{\mu_{s}+\mu_{p}}=0.1,  &  \lambda_{\mu}&=\frac{\mu_{air}}{\mu_{d}}=0.017\\  \lambda_{\rho}&=\frac{\rho_{air}}{\rho_{d}}=0.00119.
\end{align*}

The computations have been performed for 100 time steps, and the corresponding wall-clock times are averaged over this interval. 
Fig. \ref{fig:figdual5}(a) shows the wall-clock time versus the number of cores for the Newtonian case; we see that a speed-up of approximately two  is achieved 
when comparing the single-resolution solution with the same grid for the phase field parameter. 
Fig. \ref{fig:figdual5} (b) presents a  speed-up larger than a factor 2  (more than 3 for a small number of cores)  by using the dual-resolution for the viscoelastic case even if just 3.8 \% of the computational domain is occupied by the non-Newtonian phase. It is noteworthy that we obtain a gain in computational time with same accuracy as the single-resolution results.    

\section{Conclusions}\label{sec5}

We present an efficient dual-resolution approach for the numerical solution of viscoelastic two-phase flow problems, based on the Cahn-Hilliard phase-field model incorporating a dynamic contact line. The method uses the finite difference method for the discretization of the governing equations, and it is second-order in both time and space. The code has an efficient parallelization and the Helmholtz and Poisson equations are solved by a direct FFT-based solver for the two different problem sizes \citep{Li2010}. The implementation is validated against several benchmark cases with and without contact line in 2D and 3D, and the results show a very good agreement for all cases. Our investigation reveals that around 4 to 5 grid points at the nominal interface are enough to resolve the velocity field, pressure, and polymer stresses on the dual-resolution grid without affecting the accuracy of the simulation, as long as the phase field variable is solved on the twice-refined grid. For the test cases presented here, around 35\% of the computational time is spent on solving the polymeric stresses. The results show a speed-up of a factor 2 or more when just 3.8 \% of the domain is occupied by the non-Newtonian phase; a larger speed-up is expected as the non-Newtonian phase occupies a larger portion of the computational domain since the six additional equations for the polymeric stresses which need to be solved in the non-Newtonian phase by using the computationally expensive log-conformation method can be resolved on the coarser grid.

In this work we have considered wetting and low Reynolds number viscoelastic flows. However, we expect the same two-phase flow solver to be efficient in simulating also elastoviscoplastic droplet-laden flows \citep{Izbassarov2020}, and non-Newtonian turbulent two-phase flows, both of which are left to future work.

\section*{Acknowledgments}

This project has received funding from the European Research Council (ERC) Starting Grant No. 852529 under the European Union’s Horizon 2020 research and innovation programme (StG MUCUS, No. 852529). OT and KB also acknowledge the financial support by the Swedish Research Council through Grant No. VR2017-4809. 
LB thanks the Swedish Research Council, via the multidisciplinary research environment INTERFACE (VR 2016-06119 "Hybrid multiscale modelling of transport phenomena for energy efficient processes").We acknowledge the computing time on the supercomputer Beskow at the PDC center, KTH provided by SNIC (Swedish National Infrastructure for Computing),Sweden.




\bibliography{wileyNJD-AMA}%


\end{document}